\documentclass[11pt]{article}
\usepackage{axodraw}
\usepackage{graphicx}
\usepackage{amssymb}
\RequirePackage{mathptmx}
\RequirePackage[T1]{fontenc}
\usepackage{heppennames2}
\usepackage{lhchiggs}
\usepackage{cernunits}
\DeclareSymbolFont{forjmath}{OT1}{cmr}{m}{sl}
\DeclareMathSymbol{\Jmath}{\mathord}{forjmath}{'021}
\def\jmath{\Jmath}
\DeclareFontFamily{OT1}{cmr}{}
\DeclareFontFamily{OT1}{cmss}{}
\allowdisplaybreaks
\newcommand\pubdate{\today}
\def\csumb{Dipartimento di Fisica Teorica, Universit\`a di Torino, Italy\\
           INFN, Sezione di Torino, Italy}
\def\support{\footnote{Work supported by MIUR under contract
    2001023713$\_$006 and by Compagnia di San Paolo under contract ORTO11TPXK.}}
\def\Title#1{\begin{center} {\Large\bf #1 } \end{center}}

\def\Author#1{\begin{center}{ \sc #1} \end{center}}
\def\Address#1{\begin{center}{ 
\normalsize \bfseries \itshape #1} \end{center}}

\newcommand\pubblock{\rightline{\begin{tabular}{l} \\
         \pubdate\\  \end{tabular}}}
\newenvironment{Abstract}{\begin{quotation}  }{\end{quotation}}

\def\Acknowledgments{\bigskip  \bigskip \begin{center}
          \large\bf Acknowledgments\end{center}}
\def\email#1{\footnote{\tt EMAIL: #1}}
\makeatletter
\renewcommand{\@makecaption}[2]{%
   \vskip 10pt
   \setbox\@tempboxa\hbox{\small #1: #2}
   \ifdim \wd\@tempboxa >\hsize     % IF longer than one line:
       \small #1: #2\par          %   THEN set as ordinary paragraph.
     \else                        %   ELSE  center.
       \hbox to\hsize{\hfil\box\@tempboxa\hfil}
   \fi}
 \def\citenum#1{{\def\@cite##1##2{##1}\cite{#1}}}
\def\citea#1{\@cite{#1}{}}
\newcount\@tempcntc
\def\@citex[#1]#2{\if@filesw\immediate\write\@auxout{\string\citation{#2}}\fi
  \@tempcnta\z@\@tempcntb\m@ne\def\@citea{}\@cite{\@for\@citeb:=#2\do
    {\@ifundefined
       {b@\@citeb}{\@citeo\@tempcntb\m@ne\@citea\def\@citea{,}{\bf }\@warning
       {Citation `\@citeb' on page \thepage \space undefined}}%
    {\setbox\z@\hbox{\global\@tempcntc0\csname b@\@citeb\endcsname\relax}%
     \ifnum\@tempcntc=\z@ \@citeo\@tempcntb\m@ne
       \@citea\def\@citea{,}\hbox{\csname b@\@citeb\endcsname}%
     \else
      \advance\@tempcntb\@ne
      \ifnum\@tempcntb=\@tempcntc
      \else\advance\@tempcntb\m@ne\@citeo
      \@tempcnta\@tempcntc\@tempcntb\@tempcntc\fi\fi}}\@citeo}{#1}}
\def\@citeo{\ifnum\@tempcnta>\@tempcntb\else\@citea\def\@citea{,}%
  \ifnum\@tempcnta=\@tempcntb\the\@tempcnta\else
  {\advance\@tempcnta\@ne\ifnum\@tempcnta=\@tempcntb \else\def\@citea{--}\fi
    \advance\@tempcnta\m@ne\the\@tempcnta\@citea\the\@tempcntb}\fi\fi}
\newcommand\KW[1]{\vskip .5pt \vbox{\small Keywords:\ #1}}
\newcommand\PACS[1]{\vskip .5pt \vbox{\small PACS:\ #1}}
\newcommand\secstyle{\bfseries\raggedright}
\renewcommand\section{\@startsection{section}{1}{\z@}%
                                   {3.5ex \@plus 1.3ex \@minus .7ex}%
                                   {2.3ex \@plus.4ex \@minus .4ex}%
                                   {\normalfont\large\secstyle}}
\renewcommand\subsection{\@startsection{subsection}{2}{\z@}%
                                   {2.3ex\@plus 1ex \@minus .5ex}%
                                   {1.2ex \@plus .3ex \@minus .3ex}%
                                   {\normalfont\normalsize\secstyle}}
\renewcommand\subsubsection{\@startsection{subsubsection}{3}{\z@}%
                                   {2.3ex\@plus 1ex \@minus .5ex}%
                                   {1ex \@plus .2ex \@minus .2ex}%
                                   {\normalfont\normalsize\secstyle}}% 
\makeatother
\DeclareRobustCommand{\PA}{\HepParticle{A}{}{}\Xspace}
\DeclareRobustCommand{\PU}{\HepParticle{U}{}{}\Xspace}
\DeclareRobustCommand{\PV}{\HepParticle{V}{}{}\Xspace}
\DeclareRobustCommand{\PS}{\HepParticle{S}{}{}\Xspace}

\DeclareRobustCommand{\PX}{\HepParticle{X}{}{}\Xspace}
\DeclareRobustCommand{\PY}{\HepParticle{Y}{}{}\Xspace}
\DeclareRobustCommand{\Pf}{\HepParticle{f}{}{}\Xspace}
\DeclareRobustCommand{\PAf}{\HepAntiParticle{\Pf}{}{}\Xspace}
\DeclareRobustCommand{\PF}{\HepParticle{F}{}{}\Xspace}
\DeclareRobustCommand{\PI}{\HepParticle{I}{}{}\Xspace}
\DeclareRobustCommand{\PM}{\HepParticle{M}{}{}\Xspace}
\DeclareRobustCommand{\Ph}{\HepParticle{h}{}{}\Xspace}

\newcommand{\AC}{\rm{\scriptscriptstyle{AC}}}
\newcommand{\myLO}{\rm{\scriptscriptstyle{LO}}}
\newcommand{\myNLO}{\rm{\scriptscriptstyle{NLO}}}
\newcommand{\myYM}{\rm{\scriptscriptstyle{YM}}}
\newcommand{\myFP}{\rm{\scriptscriptstyle{FP}}}

\newcommand{\Lint}{\mathrm{int}}
\newcommand{\mySM}{\rm{\scriptscriptstyle{SM}}}
\newcommand{\MSSM}{\rm{\scriptscriptstyle{MSSM}}}
\newcommand{\BSM}{\rm{\scriptscriptstyle{BSM}}}

\newcommand{\ssF}{{\mathrm{F}}}
\newcommand{\ssR}{{\mathrm{R}}}
\newcommand{\ssD}{{\mathrm{D}}}

\newcommand{\ssL}{{\mathrm{L}}}
\newcommand{\ssY}{{\mathrm{Y}}}

\newcommand{\ssS}{{\mathrm{S}}}

\newcommand{\ssV}{{\mathrm{V}}}
\newcommand{\ssM}{{\mathrm{M}}}

\newcommand{\ssT}{{\mathrm{T}}}

\newcommand{\bqas}{\begin{eqnarray*}}
\newcommand{\eqas}{\end{eqnarray*}}
\newcommand{\nl}{\nonumber\\}

\newcommand{\lpar}{\left(}                            % bracketing
\newcommand{\rpar}{\right)}

\newcommand{\bq}{\begin{equation}}                    % equationing
\newcommand{\eq}{\end{equation}}
\newcommand{\bqa}{\arraycolsep 0.14em\begin{eqnarray}}
\newcommand{\eqa}{\end{eqnarray}}
\newcommand{\ba}[1]{\begin{array}{#1}}
\newcommand{\ea}{\end{array}}
\newcommand{\ben}{\begin{enumerate}}
\newcommand{\een}{\end{enumerate}}
\newcommand{\bei}{\begin{itemize}}
\newcommand{\eei}{\end{itemize}}
\newcommand{\eqn}[1]{Eq.(\ref{#1})}
\newcommand{\eqns}[2]{Eqs.(\ref{#1})--(\ref{#2})}

\newcommand{\appendx}[1]{Appendix~\ref{#1}}

\newcommand{\bmid}{\Bigr|}

\newcommand{\ord}[1]{{\cal O}\lpar#1\rpar}
\newcommand{\oD}{{\overline \Delta}}
\newcommand{\Bref}[1]{Ref.~\cite{#1}}
\newcommand{\Brefs}[1]{Refs.~\cite{#1}}

\newcommand{\eg}{e.g.\xspace}
\newcommand{\ie}{i.e.\xspace}
\newcommand{\etc}{etc.\@\xspace}

\newcommand{\mhs}{\mathswitch {M^2_{\PH}}}
\newcommand{\mPs}{\mathswitch {M^2_{\PS}}}

\newcommand{\mws}{\mathswitch {M^2_{\PW}}}
\newcommand{\mzs}{\mathswitch {M^2_{\PZ}}}
\newcommand{\mt}{\mathswitch {M_{\PQt}}}
\newcommand{\mb}{\mathswitch {M_{\PQb}}}
\newcommand{\mts}{\mathswitch {M^2_{\PQt}}}
\newcommand{\mbs}{\mathswitch {M^2_{\PQb}}}
\newcommand{\cph}{\mathswitch {s_{\PH}}}

\newcommand{\cpz}{\mathswitch {s_{\PZ}}}

\newcommand{\dec}{{\mbox{\scriptsize dec}}}
\newcommand{\ndec}{{\mbox{\scriptsize non-dec}}}

\newcommand{\bos}{{\mbox{\scriptsize bos}}}

\newcommand{\ren}{{\mbox{\scriptsize ren}}}

\newcommand{\eff}{{\mbox{\scriptsize eff}}}
\newcommand{\DR}{{\mbox{\scriptsize DR}}}
\newcommand{\full}{{\mbox{\scriptsize full}}}

\newcommand{\esp}{{\mbox{\scriptsize exp}}}
\newcommand{\trac}{{\mathrm{Tr}}}
\newcommand{\gfix}{{\mbox{\scriptsize gf}}}
\newcommand{\fact}{{\mbox{\scriptsize fc}}}
\newcommand{\nfact}{{\mbox{\scriptsize nfc}}}
\newcommand{\spin}{{\mbox{\scriptsize spin}}}
\newcommand{\fer}{{\mbox{\scriptsize fer}}}

\newcommand{\FP}{{\mathrm{FP}}}
\newcommand{\intfxy}[2]{\int_{\scriptstyle 0}^{\scriptstyle 1}\,d#1\,
                        \int_{\scriptstyle 0}^{\scriptstyle #1}\,d#2}
\newcommand{\li}[2]{\mathrm{Li}_{#1}\lpar\displaystyle{#2}\rpar} % polylog
\newcommand{\ep}{\mathswitch \varepsilon}

\newcommand{\spro}[2]{{#1}\cdot{#2}}

\newcommand{\stw}{s_{\theta}}             % bare, Lagrangian parameters
\newcommand{\ctw}{c_{\theta}}
\newcommand{\stws}{s_{\theta}^2}
\newcommand{\ctws}{c_{\theta}^2}
\newcommand\mybar[1]{\ensuremath{\bar{#1}}}
\newcommand\mytil[1]{\ensuremath{\tilde{#1}}}
\newcommand{\cth}{{\hat c}_{\theta}}
\newcommand{\sth}{{\hat s}_{\theta}}
\newcommand{\cths}{{\hat c}^2_{\theta}}
\newcommand{\sths}{{\hat s}^2_{\theta}}
\newcommand{\sthq}{{\hat s}^4_{\theta}}
\newcommand{\HSs}{{\Lambda^2}}
\DeclareRobustCommand{\PWpmmu}{\HepParticle{\PW}{\mu}{\pm}\Xspace}
\DeclareRobustCommand{\PWpmu}{\HepParticle{\PW}{\mu}{+}\Xspace}
\DeclareRobustCommand{\PWmmu}{\HepParticle{\PW}{\mu}{-}\Xspace}
\DeclareRobustCommand{\PWpnu}{\HepParticle{\PW}{\nu}{+}\Xspace}
\DeclareRobustCommand{\PWmnu}{\HepParticle{\PW}{\nu}{-}\Xspace}
\DeclareRobustCommand{\PZmu}{\HepParticle{\PZ}{\mu}{}\Xspace}

\DeclareRobustCommand{\PXpm}{\HepParticle{\PX}{}{\pm}\Xspace}
\DeclareRobustCommand{\PXp}{\HepParticle{\PX}{}{+}\Xspace}
\DeclareRobustCommand{\PXm}{\HepParticle{\PX}{}{-}\Xspace}
\DeclareRobustCommand{\PYz}{\HepParticle{\PY}{Z}{}\Xspace}
\DeclareRobustCommand{\PYa}{\HepParticle{\PY}{A}{}\Xspace}
\DeclareRobustCommand{\PAXpm}{\HepAntiParticle{\PX}{}{\,\pm}\Xspace}
\DeclareRobustCommand{\PAXp}{\HepAntiParticle{\PX}{}{\,+}\Xspace}
\DeclareRobustCommand{\PAXm}{\HepAntiParticle{\PX}{}{\,-}\Xspace}
\DeclareRobustCommand{\PAYz}{\HepAntiParticle{\PY}{\,\PZ}{}\Xspace}
\DeclareRobustCommand{\PAYa}{\HepAntiParticle{\PY}{\,\PA}{}\Xspace}
\newcommand{\POZ}{{\overline{\PZ}}}
\newcommand{\POA}{{\overline{\PA}}}
\newcommand{\POH}{{\overline{\PH}}}

\DeclareRobustCommand{\POppm}{\HepAntiParticle{\upphi}{}{\,\pm}\Xspace}
\DeclareRobustCommand{\POpp}{\HepAntiParticle{\upphi}{}{\,+}\Xspace}
\DeclareRobustCommand{\POpm}{\HepAntiParticle{\upphi}{}{\,-}\Xspace}
\DeclareRobustCommand{\POpz}{\HepAntiParticle{\upphi}{}{\,0}\Xspace}
\DeclareRobustCommand{\Ppp}{\HepParticle{\upphi}{}{+}\Xspace}
\DeclareRobustCommand{\Ppm}{\HepParticle{\upphi}{}{-}\Xspace}
\DeclareRobustCommand{\Ppz}{\HepParticle{\upphi}{}{0}\Xspace}
\DeclareRobustCommand{\PpsiL}{\HepParticle{\uppsi}{L}{}\Xspace}
\DeclareRobustCommand{\PpsiR}{\HepParticle{\uppsi}{R}{}\Xspace}
\DeclareRobustCommand{\POpsiL}{\HepParticle{\mybar{\uppsi}}{L}{}\Xspace}
\DeclareRobustCommand{\POpsiR}{\HepParticle{\mybar{\uppsi}}{R}{}\Xspace}
\DeclareRobustCommand{\PQtR}{\HepParticle{\PQt}{R}{}\Xspace}
\DeclareRobustCommand{\PQbR}{\HepParticle{\PQb}{R}{}\Xspace}
\newcommand{\pdmu}{{\partial_{\mu}}}
\newcommand{\pdnu}{{\partial_{\nu}}}
\newcommand{\myGF}{G_{\ssF}}
\DeclareRobustCommand{\PK}{\HepParticle{K}{}{}\Xspace}
\DeclareRobustCommand{\PKdag}{\HepParticle{\PK}{}{\dagger}\Xspace}
\newcommand{\KdK}{\lpar \PKdag\,\PK\rpar }
\DeclareRobustCommand{\Pxi}{\HepParticle{\upxi}{}{}\Xspace}
\DeclareRobustCommand{\Pxip}{\HepParticle{\upxi}{}{+}\Xspace}
\DeclareRobustCommand{\Pxim}{\HepParticle{\upxi}{}{-}\Xspace}
\DeclareRobustCommand{\Pxiz}{\HepParticle{\upxi}{}{0}\Xspace}
\DeclareRobustCommand{\Pxid}{\HepParticle{\upxi}{}{\dagger}\Xspace}
\newcommand{\dPK}{{\partial\PK}}
\newcommand{\essV}{{\mathrm{e}\PV}}
\DeclareRobustCommand{\PAf}{\HepAntiParticle{\Pf}{}{}\Xspace}
\newcommand{\Lag}{{\cal L}}
\newcommand{\Ope}{{\cal O}}
\newcommand{\Amp}{{\cal M}}
\DeclareRobustCommand{\xiW}{\HepParticle{\upxi}{\PW}{}\Xspace}
\DeclareRobustCommand{\xiZ}{\HepParticle{\upxi}{\PZ}{}\Xspace}
\DeclareRobustCommand{\OxiZ}{\HepParticle{\mybar{\upxi}}{\PZ}{}\Xspace}
\DeclareRobustCommand{\OxiZs}{\HepParticle{\mybar{\upxi}}{\PZ}{2}\Xspace}
\DeclareRobustCommand{\PWpL}{\HepParticle{\PW}{\PL}{+}\Xspace}
\DeclareRobustCommand{\PWmL}{\HepParticle{\PW}{\PL}{-}\Xspace}
\newcommand{\Fsc}{\mu^2_{\ssF}}
\newcommand{\Rsc}{\mu^2_{\ssR}}
\DeclareRobustCommand{\Ph}{\HepParticle{h}{}{}\Xspace}
\DeclareRobustCommand{\PHpm}{\HepParticle{\PH}{}{\pm}\Xspace}
\DeclareRobustCommand{\PHp}{\HepParticle{\PH}{}{+}\Xspace}
\DeclareRobustCommand{\PHm}{\HepParticle{\PH}{}{-}\Xspace}
\DeclareRobustCommand{\Pft}{\HepParticle{\tilde{\Pf}}{}{}\Xspace}
\DeclareRobustCommand{\Pcpmi}{\HepParticle{\upchi}{i}{\pm}\Xspace}
\DeclareRobustCommand{\Pcpi}{\HepParticle{\upchi}{i}{+}\Xspace}
\DeclareRobustCommand{\Pcmi}{\HepParticle{\upchi}{i}{-}\Xspace}
\DeclareRobustCommand{\Pci}{\HepParticle{\upchi}{i}{}\Xspace}
\newcommand{\srt}{\sqrt{2}}
\newcommand{\PWW}{\PW\PW}
\newcommand{\PZZ}{\PZ\PZ}
\DeclareRobustCommand{\PdT}{\HepParticle{\mytil{\PF}}{\mu\nu}{a}\Xspace}
\DeclareRobustCommand{\HB}{\HepAntiParticle{\PH}{}{}\Xspace}
\DeclareRobustCommand{\PZB}{\HepAntiParticle{\PZ}{}{}\Xspace}
\DeclareRobustCommand{\myPAB}{\HepAntiParticle{\PA}{}{}\Xspace}
\DeclareRobustCommand{\OM}{\HepAntiParticle{\PM}{}{}\Xspace}
\DeclareRobustCommand{\OMs}{\HepAntiParticle{\PM}{}{\,2}\Xspace}
\DeclareRobustCommand{\OMzs}{\HepAntiParticle{\PM}{\,0}{\,2}\Xspace}
\DeclareRobustCommand{\OMzq}{\HepAntiParticle{\PM}{\,0}{\,4}\Xspace}
\DeclareRobustCommand{\OMHs}{\HepAntiParticle{\PM}{\,\PH}{\,2}\Xspace}
\DeclareRobustCommand{\OMHq}{\HepAntiParticle{\PM}{\,\PH}{\,4}\Xspace}
\DeclareRobustCommand{\phib}{\HepAntiParticle{\upphi}{}{}\Xspace}
\newcommand{\ACf}{\frac{\OMs}{\HSs}}
\DeclareRobustCommand{\PHb}{\HepParticle{\mathbf{H}}{}{}\Xspace}
\DeclareRobustCommand{\PAb}{\HepParticle{\mathbf{A}}{}{}\Xspace}
\DeclareRobustCommand{\PZb}{\HepParticle{\mathbf{Z}}{}{}\Xspace}
\DeclareRobustCommand{\PWb}{\HepParticle{\mathbf{W}}{}{}\Xspace}
\DeclareRobustCommand{\PWpb}{\HepParticle{\mathbf{W}}{}{+}\Xspace}
\DeclareRobustCommand{\PWmb}{\HepParticle{\mathbf{W}}{}{-}\Xspace}
\DeclareRobustCommand{\Pgb}{\HepParticle{\mathbf{g}}{}{}\Xspace}
\DeclareRobustCommand{\PGgb}{\HepParticle{\mathbf{\mathbf{\upgamma}}}{}{}\Xspace}
\DeclareRobustCommand{\PQtb}{\HepParticle{\mathbf{t}}{}{}\Xspace}
\DeclareRobustCommand{\PAQtb}{\HepAntiParticle{\mathbf{t}}{}{}\Xspace}
\DeclareRobustCommand{\PQbb}{\HepParticle{\mathbf{b}}{}{}\Xspace}
\DeclareRobustCommand{\PAQbb}{\HepAntiParticle{\mathbf{b}}{}{}\Xspace}
\DeclareRobustCommand{\Pfb}{\HepParticle{\mathbf{f}}{}{}\Xspace}
\DeclareRobustCommand{\PAfb}{\HepAntiParticle{\mathbf{f}}{}{}\Xspace}
\begin{document}
\begin{titlepage}
\pubblock
%\pubdate
%
\vfill
\def\thefootnote{\fnsymbol{footnote}}
\Title{\LARGE \sffamily \bfseries
NLO Inspired Effective Lagrangians\\[0.2cm]
for Higgs Physics\support
}
\vfill
\Author{\normalsize \bfseries \sffamily
Giampiero Passarino \email{giampiero@to.infn.it}}               
\Address{\csumb}
\vfill
\vfill
\begin{Abstract}
\noindent 
Either late autumn this year or latest early next year LHC should have results with 
$2{-}3$ times the current data which migth give first clues on the couplings
of the light narrow resonance. A strategy for measuring deviations from the
Standard Model can be based on using the ``full'' Standard Model, including all available QCD and
electroweak higher-order corrections, and supplement it with $d= 6$ local operators. Their
Wilson coefficients are assumed to be small enough that they can be treated at
leading order. Examples of the connection of local operators with BSM Lagrangians are
presented as well as a discussion of Lagrangians with/without decoupling of heavy degrees of
freedom. The whole strategy is critically reviewed in the light of internal consistency.
\end{Abstract}
\vfill
\begin{center}
\KW{Feynman diagrams, loop calculations, radiative corrections, effective Lagrangian,
Higgs physics} 
%Keywords: Feynman diagrams, loop calculations, radiative corrections, effective Lagrangian,
%Higgs physics \\[5mm]
\PACS{11.15.Bt, 12.38.Bx, 13.85.Lg, 14.80.Bn, 14.80.Cp}
%PACS classification: 11.15.Bt, 12.38.Bx, 13.85.Lg, 14.80.Bn, 14.80.Cp
\end{center}
\end{titlepage}
\def\thefootnote{\arabic{footnote}}
\setcounter{footnote}{0}
%--
\small
\thispagestyle{empty}
\tableofcontents
\normalsize
%--
\clearpage
%--
\setcounter{page}{1}
%--
\section{Introduction}
%--
An interesting question is how present and future experiments will be able to probe the
couplings of the Higgs boson at a high level of precision, see \Bref{Peskin:2012we} for
a discussion. There is a wide variety of beyond the Standard Model (BSM) theories where 
the Higgs couplings differs from the Standard Model (SM) ones by less that $10\%$, as discussed 
in \Bref{Gupta:2012mi}.
Among many papers dealing with the subject we quote those in 
\Brefs{Zeppenfeld:2000td,Belyaev:2002ua,Duhrssen:2004cv,Lafaye:2009vr,Klute:2012pu}. 
For the most recent update on the subject we refer to the work of 
\Brefs{Espinosa:2012vu,Espinosa:2012im,Espinosa:2012ir}.
Interim recommendations to explore the coupling structure of a Higgs-like particle
can be found in \Bref{LHCHiggsCrossSectionWorkingGroup:2012np}.

In this work, following Independence Day~\cite{:2012gk,:2012gu}, we imagine that there is 
a huge space of theories, represented by local and renormalizable Lagrangians where 
SM is only one point.  A possible strategy to look for deviations from the 
SM is the following:
%--
\bei

\item we take the SM as the theory of ``light'' degrees of freedom, \ie $d= 4$
  operators

\item we simulate the unknown extension of the SM by the most general set of
  $d= 6$ operators that are obtained by integrating out the heavy degrees of
  freedom (we also assume no sensitivity to operators with $d \ge 8$ at LHC).
  This is equivalent to say that the BSM theory is unknown or matching is too
  difficult to carry out, so we write the most general set of interactions 
  consistent with symmetries. The effective theory contains an infinite number
  of operators but only a finite number is needed for present (LHC) precision.

\eei
%--
With enough statistics it should be possible to fit $a_i$, the Wilson coefficients of the $d= 6$
operators, and there are two possibilities: a) they are close to zero (where zero = SM) or 
b) they are not.
Option a) tell us that NLO corrections (or the residual theoretical uncertainty at
NNLO level) and the $a_i$ coefficients are small, and the SM is actually a minimum in our 
Lagrangian space or very close to it. This will explain nothing but it is internally consistent.
Option b) raises serious problems since the effect of local operators is large and they cannot be
included only at LO, but inserting operators in SM loops creates even more problems.

In case it is option b) we should move in the Lagrangian space and adopt a new renormalizable 
Lagrangian with the virtue of making zero that specific (large) Wilson coefficient $a_i$;
local operators are then redefined w.r.t. the new Lagrangian. Of course there will be more 
Lagrangians projecting into the same set of operators but still we could see how our new 
choice handles the rest of the data.

In principle, there will be a blurred arrow in our space of Lagrangians, and we should simply 
focus the arrow. Without invoking the explicit example of Supersymmetry this
is the so-called inverse problem introduced in \Brefs{ArkaniHamed:2005px,ArkaniHamed:2007fw}:
if LHC finds evidence for physics beyond the SM, how can one determine the underlying theory?

It is worth noting that this question is highly difficult to receive a complete answer at 
the LHC. The main goal will be to identify the structure of the effective Lagrangian 
(\ie the different scalings of the various $d=6$ operators) and to derive qualitative
information on new physics; the question of the ultraviolet (UV) completion cannot be answered
unless there is sensitivity to $d > 6$ operators. Therefore, we are looking for a relatively
modest goal on the road to understand if the effective theory can be UV completed 
(bottom-up approach with no obvious embedding).  

To set up our definitions of an enlarged theory we have to specify the concept of Higgs fields:
it is the set of scalar fields that break electroweak symmetry (EW) by developing a vacuum
expectation value (VEV). What we are looking for is evidence of SM Higgs properties or
deviations from the SM behavior; in the latter case one has to understand consistency with other 
EW symmetry-breaking frameworks. Alternatively we can consider scenarios with more scalar fields, 
that are not Higgs fields (Higgs partners); the problem with more VEVs, or one VEV
different from $(T\,,\,Y)= (\frac{1}{2}\,,\,1)$ ($T$ is isospin and $Y$ is hypercharge), is 
partially related to the rho-parameter~\cite{Ross:1975fq} which at tree-level is given by
%--
\bq
\uprho_{\myLO} = \frac{1}{2}\,\frac{\sum_i\,\Bigl[ c_i\,\mid v_i\mid^2 + r_i\,u^2_i\Bigr]}
{\sum_i\,Y^2_i\,\mid v_i\mid^2},
\qquad 
c_i = T_i\,\lpar T_i + 1\rpar - Y^2_i,
\quad
r_i = T_i\,\lpar T_i + 1\rpar,
\eq
%--
where the sum is over all Higgs fields, $v_i(u_i)$ gives the VEV of a complex(real) Higgs 
field with hypercharge $Y_i$ and weak-isospin $T_i$. Our considerations will be presented in 
\refS{acdec}. The experimental limit on $\uprho - 1$ are rather stringent.
For a complete discussion of models respecting custodial symmetry we refer to \Bref{Low:2010jp}. 

In this paper we do not discuss questions related to spin, mass or CP quantum numbers but
only couplings. In particular we discuss couplings to vector bosons since they control the 
unitarity behavior of longitudinal $\PV\PV\,$-scattering at high energy~\cite{Passarino:1990hk}
(automatic in the SM). We also discuss the effects of Higgs partners and of Higgs self-couplings 
in \refS{dHp}.

For a better illustration of our approach we observe that a consistent effective theory, defined by
%--
\bq
\Lag= \Lag_4 + \sum_{n > 4}\,\sum_{i=1}^{N_n}\,\frac{a^n_i}{\Lambda^{n-4}}\,\Ope^{(d=n)}_i
\eq
%--
has arbitrary Wilson coefficients $a^n_i$ which, however, give the leading amplitudes in an exactly
unitary $S\,$-matrix at energies far below $\Lambda$. The theory is non-renormalizable, which
means that an infinite number of higher operators must be included. Nevertheless there is a 
consistent expansion of amplitudes in power of $E/\Lambda$. Our goal will be to understand the 
$d= 6$ operators as a first step towards an UV completion, possibly a weakly-coupled one \ie
one where weakly-coupled new physics opens up around $\Lambda$ and restores unitarity 
(a different scenario, {\em classicalization}, has been proposed in 
\Brefs{Dvali:2010jz,Dvali:2010ns}).

In other words, the question is: can we classify the low-energy (LHC) observables that
determine the road to UV completion?
Note that there is a claim in the literature~\cite{Adams:2006sv,Dvali:2012zc} that the 
coefficients $a_i$ must be positive to have an UV completion which respects the usual axioms of 
$S\,$-matrix theory.
In particular, in the work of \Bref{Dvali:2012zc} it is shown that UV completion is
encoded in the sign of the scattering amplitude for longitudinal vector-bosons and that
weakly-coupled UV completion requires a positive sign.
Constraints on the sign of the couplings in an effective Higgs Lagrangian using prime principles
have been derived in \Bref{Low:2009di}.

In \refS{Def}, we present the SM Lagrangian, and in \refS{AC} we introduce the effective 
Lagrangian. We discuss Higgs vertices in \refS{aHv} and $\PZ$ vertices in \refS{aZv}.
\refS{pdw} gives the relevant partial decay widths of the $\PH$ boson. In \refS{4fdec} we list 
the various $\PH \to 4\,\Pf$ decays. Double Higgs production is discussed in \refS{dHp}.
We discuss perturbative unitarity in \refS{pu}. We give our conclusions in \refS{conclu}.
%--
\section{$\Lag_{\mySM}$: definitions \label{Def}}
%--
In this Section we collect all definitions that are needed to write the SM 
Lagrangian~\cite{Bardin:1999ak}. The scalar field $\PK$ (with hypercharge $1/2$) is defined by
%--
\[
\PK = \frac{1}{\srt}\,\left(
\begin{array}{c}
\PH + 2\,\frac{M}{g} + i\,\Ppz \\
\srt\,i\,\Ppm
\end{array}
\right)
\]
%--
$\PH$ is the custodial singlet in $\lpar 2_{\ssL}\,\otimes\,2_{\ssR}\rpar = 1\,\oplus\,3$.
Charge conjugation gives $K^c_i = \ep_{ij}\,K^*_j$, or
%--
\[
\PK^c = -\,\frac{1}{\srt}\,\left(
\begin{array}{c}
\srt\,i\,\Ppp \\
\PH + 2\,\frac{M}{g} - i\,\Ppz 
\end{array}
\right)
\]
%--
The covariant derivative $D_{\mu}$ is 
%--
\bq
D_{\mu}\,\PK = \Bigl( \pdmu - \frac{i}{2}\,g_0\,B^a_{\mu}\,\uptau_a - 
\frac{i}{2}\,g\,g_1\,B^0_{\mu}\Bigr)\,\PK
\eq
%--
with $g_1 = -\stw/\ctw$ and where $\uptau^a$ are Pauli matrices while $\stw(\ctw)$ is the
sine(cosine) of the weak-mixing angle. Furthermore
%--
\bq
\PWpmmu = \frac{1}{\srt}\,\lpar B^1_{\mu} \mp i\,B^2_{\mu}\rpar ,
\qquad
\PZmu = \ctw\,B^3_{\mu} - \stw\,B^0_{\mu},
\quad
\PA_{\mu} = \stw\,B^3_{\mu} + \ctw\,B^0_{\mu}.
\eq
%--
\bq
F^a_{\mu\nu} = \pdmu\,B^a_{\nu} - \pdnu\,B^a_{\mu}
+ g_0\,\epsilon^{a b c}\,B^b_{\mu}\,b^c_{\nu},
\quad
F^0_{\mu\nu} = \pdmu\,B^0_{\nu} - \pdnu\,B^0_{\mu}.
\eq
%--
Here $a,b,\dots = 1,\dots,3$. The dual tensor is defined by
%--
\bq
\PdT = \ep^{\mu\nu\alpha\beta}\,\PF^a_{\alpha\beta}.
\eq
%--
Furthermore, for the QCD part we introduce
%--
\bq
G^a_{\mu\nu} = \pdmu\,g^a_{\nu} - \pdnu\,g^a_{\mu}
+ g_{\ssS}\,f^{a b c}\,g^b_{\mu}\,g^c_{\nu}.
\eq
%--
Here $a,b,\dots = 1,\dots,8$ and the $f$ are the $SU(3)$ structure constants. 
Finally, we introduce fermions,
%--
\[
\PpsiL = \left(
\begin{array}{c}
\PQt \\ \PQb
\end{array}
\right)_{\ssL}
\qquad
\Pf_{\ssL\,\ssR} = \frac{1}{2}\,\lpar 1 \pm \gamma^5\rpar \,\Pf
\]
%--
and their covariant derivatives
%--

%--
\bqa
D_{\mu}\,\PpsiL &=& \lpar \pdmu + g\,B^i_{\mu}\,T_i\rpar \,\PpsiL,
\quad i=0,\dots,3
\nl
T^a &=& -\frac{i}{2}\,\uptau^a,  \qquad
T^0 = -\frac{i}{2},g_2\,I,
\eqa
%--
\bq
D_{\mu}\,\PpsiR = \lpar \pdmu + g\,B^i_{\mu}\,t_i\rpar \,\PpsiR,
\qquad t^a = 0,
\eq
%--
\[
t^0 = -\frac{i}{2}\,\left(
\begin{array}{cc}
g_3 & 0 \\
0 & g_4
\end{array}
\right)
\]
%--
with $g_i= -\stw/\ctw\,\uplambda_i$ and
%--
\bq
\uplambda_2 = 1 - 2\,Q_u, 
\quad
\uplambda_3 =  - 2\,Q_u, 
\quad
\uplambda_4 =  - 2\,Q_d.
\eq 
%--
The Standard Model Lagrangian is the sum of several terms:
%--
\bq
\Lag_{\mySM} = \Lag_{\myYM} + \Lag_{\PK} + \Lag{\gfix} +
\Lag_{\myFP} + \Lag_{\Pf}
\eq
%--
\ie, Yang-Mills, scalar, gauge-fixing, Faddeev-Popov ghosts and fermions.
Furthermore, for a proper treatment of the neutral sector of the SM, we introduce a new
coupling constant $g$, defined by the relation
%--
\bq
g_0 = g\,\lpar 1 + g^2\,\Gamma \rpar,
\eq
%--
where $\Gamma$ is fixed by the request that the $\PZ-\PA$ transition is zero at $p^2= 0$,
see \Bref{Actis:2006ra}.
%--
The scalar Lagrangian is given by
%--
\bq
\Lag_{\PK} = - \,\lpar D_{\mu}\,\PK\rpar^{\dagger}\,D_{\mu}\,\PK -
\mu^2\,\PKdag\,\PK - \frac{1}{2}\,\uplambda\,\lpar \PKdag\,\PK\rpar^2.
\eq
%--
We will work in the $\beta_h\,$-scheme~\cite{Actis:2006ra}, where
%--
\bq
\mu^2 = \beta_h - 2\,\frac{\uplambda}{g}\,M^2,
\qquad
\uplambda = \frac{1}{4}\,g^2\,\frac{\mhs}{M^2}
\eq
%--
Furthermore, we introduce $v= \srt\,M/g$.
%--
and fix $\beta_h$ order-by-order in perturbation theory by requiring $<\,0\,|\,\PH\,|\,0\,> = 0$.
%--
\section{Simplified effective Lagrangian \label{AC}}
%--
Our minimal list of $d= 6$ operators is based on the work of 
\Brefs{Buchmuller:1985jz,Grzadkowski:2010es} 
and of \Brefs{Bonnet:2012nm,Bonnet:2011yx,Kanemura:2008ub,Horejsi:2004fs} (see also 
\Brefs{Hagiwara:1993qt,Hankele:2006ma}, 
\Bref{Anastasiou:2011pi},
\Brefs{Corbett:2012dm,Qi:2008ex,Hasegawa:2012mf,Degrande:2012gr,Azatov:2012rd},
\Brefs{GonzalezGarcia:1999fq,Eboli:1998vg,Barger:2003rs} and \Bref{delAguila:2010mx})
and is given in \refT{d6list}.
%--
\bq
\Lag = \Lag_{\mySM} + \sum_i\,\frac{a_i}{\HSs}\,\Ope^{d=6}_i,
\label{FLag}
\eq
%--
%where
%--
\begin{table}
\begin{center}
\caption[]{\label{d6list}{A selection of relevant $d = 6$ operators}}
\vspace{0.2cm}
\begin{tabular}{ll}
\hline
& \\
$\Ope_{\PK} = -\frac{g^3}{3}\,\KdK^3 $&
$\Ope_{\dPK} = \frac{g^2}{2}\,\pdmu\,\KdK\,\pdmu\,\KdK $\\
$\Ope^1_{\PK} = g^2\,\KdK\,\lpar D_{\mu}\,\PK\rpar ^\dagger\,D_{\mu}\,\PK $&
$\Ope^3_{\PK} = g^2\,\lpar \PKdag D_{\mu} \PK\rpar\,\Bigl[\lpar D_{\mu} \PK\rpar^\dagger \PK\Bigr] $\\
$\Ope^4_{\PK} = i\,g^2\,\lpar D_{\mu} \PK\rpar ^\dagger\,\uptau_a\,D_{\mu} \PK\,F^a_{\mu\nu} $&
$\Ope^5_{\PK} = i\,g^2\,\lpar D_{\mu} \PK\rpar ^\dagger\,D_{\mu} \PK\,F^0_{\mu\nu} $\\
$\Ope^1_{\PV} = g\,\lpar  \PKdag\,\PK - v^2\rpar \,F^a_{\mu\nu}\,F^a_{\mu\nu} $&
$\Ope^2_{\PV} = g\,\lpar  \PKdag\,\PK - v^2\rpar \,F^0_{\mu\nu}\,F^0_{\mu\nu} $\\
$\Ope^3_{\PV} = g\,\PKdag\,\uptau_a\,\PK\,F^a_{\mu\nu}\,F^0_{\mu\nu} $&
$\Ope^1_{{\essV}} = g\,\lpar  \PKdag\,\PK - v^2\rpar \,\PdT\,\PF^a_{\mu\nu} $\\
$\Ope^2_{{\essV}} = g\,\lpar  \PKdag\,\PK - v^2\rpar \,\PF^0_{\mu\nu}\,\PdT $&
$\Ope^3_{{\essV}} = g\,\PKdag\,\uptau_a\,\PK\,\PdT\,F^0_{\mu\nu} $\\
$\Ope_{\Pg} = g\,\lpar  \PKdag\,\PK - v^2\rpar \,G^a_{\mu\nu}\,G^a_{\mu\nu} $&
$\Ope^1_{\Pf} = g^2\,\lpar  \PKdag\,\PK - v^2\rpar \,\POpsiL\,\PK\,\PQtR + \mbox{h. c.} $\\
$\Ope^2_{\Pf} = g^2\,\lpar  \PKdag\,\PK - v^2\rpar \,\POpsiL\,\PK^c\,\PQbR + \mbox{h. c.} $&
$\Ope^3_{\Pf} = \POpsiL\,D_{\mu}\,\PQtR\,D_{\mu}\,\PK + \mbox{h. c.} $\\
$\Ope^4_{\Pf} = \POpsiL\,D_{\mu}\,\PQbR\,D_{\mu}\,\PK^c + \mbox{h. c.}$ & \\
& \\
\hline
\end{tabular}
\end{center}
\end{table}
%--

The structure of the $d = 6$ operators is chosen in such a way that, with $\beta_h = 0$,
no term proportional to $1/g$ will appear in the Lagrangian (a part from irrelevant constant
terms). Operators containing $\PdT$ are CP-odd, the remaining ones are CP-even. 

Additional operators not included in \refT{d6list} have been considered in Eq. (4) of
\Bref{Degrande:2012gr} and are given in \refT{d6alist}.
%--
\begin{table}
\begin{center}
\caption[]{\label{d6alist}{Alternative single-fermionic-current $d = 6$ operators}}
\vspace{0.2cm}
\begin{tabular}{l}
\hline
 \\
$\Ope^5_{\Pf} = \lpar \PK^{\dagger} D_{\mu} \PK\rpar\,\POpsiL \gamma^{\mu} \PpsiL + \mbox{h. c.}
$ \\
$\Ope^6_{\Pf} = \lpar \PK^{\dagger} D_{\mu} \PK\rpar\,\POpsiR \gamma^{\mu} \PpsiR + \mbox{h. c.}
$ \\
$\Ope^7_{\Pf} = \lpar \PK^{\dagger} \uptau_a D_{\mu} \PK\rpar\,
                \POpsiL \uptau^a \gamma^{\mu} \PpsiL + \mbox{h. c.} $ \\
 \\
\hline
\end{tabular}
\end{center}
\end{table}
%--
In certain models their effect can be comparable to the one of $\Ope_{\Pg}$.
However, they do not contribute to the $\PH \PAQq \PQq$ vertex, as explained in
\Bref{Degrande:2012gr}, because the vector current $\PAQq \gamma^{\mu} \PQq$ is conserved.
For a complete list of $d = 6$ operators (other than the four-fermion ones) we refer to
Table~2 of \Bref{Grzadkowski:2010es}.

For the single-fermionic-current operators we have adopted the (simplified) choice of 
\Bref{Kanemura:2008ub}, discarding the chromomagnetic dipole moment operator, which affects 
the process $\Pg\Pg \to \PAQt \PQt$; in general it is known how to remove derivatives acting 
on the spinors using integration-by-parts. For a complete classification we refer again to 
Table~2 of \Bref{Grzadkowski:2010es} where there are $13$ operators of dimension six involving 
single-currents of quark fields. 

If one restricts the analysis to the calculation of on-shell matrix elements then
there are linear combinations of operators that vanish by the Equations-Of-Motion (EOM).
Under this assumption there are redundant operators, \eg $\Ope^1_{\PK}$, which
can be expressed in terms of a $d = 4$ operator $\KdK^2$, of $\Ope_{\PK}, \Ope_{\dPK}$ and 
of higher dimensional Yukawa interactions involving $\mybar{\psi}\,\psi$ and three $K\,$-fields, 
\ie $\Ope^{1,2}_{\Pf}$.
Since we are working with unstable particles, the use of EOM should be taken with due
caution; indeed, only $S\,$-matrix elements will be the same for equivalent operators but not
the Green's functions. 

It has been pointed out in \Bref{Wudka:1994ny} that, even if the 
$S\,$-matrix elements cannot distinguish between two equivalent operators ${\cal O}$ and
${\cal O}'$, there is a large quantitative difference whether the underlying theory can generate
${\cal O}'$ or not. It is equally reasonable not to eliminate redundant operators and,
eventually, exploit redundancy to check $S\,$-matrix elements.
If one eliminates them, whenever the Higgs boson is taken on-shell and the full 
set of $d = 6$ operators of \Bref{Grzadkowski:2010es} is used, the presence of $\Ope^1_{\PK}$ 
is redundant, and one should set $a^1_{\PK} = 0$. Strictly speaking, the last statement only 
applies to single-Higgs processes; the argument is simple (see Appendix.~D of \Bref{Wudka:1994ny}),
given a theory with a Lagrangian $\Lag\,[\upphi]$ consider an effective Lagrangian 
$\Lag_{\eff} = \Lag + g\,\Ope + g'\,\Ope'$ where 
$\Ope - \Ope' = F[\upphi]\,\delta\Lag/\delta\upphi$, and $F$ is some local functional of
$\upphi$. The effect of $\Ope'$ on $\Lag_{\eff} = \Lag + g\,\Ope$ is to shift
$g \to g + g'$ and to replace $\upphi \to \upphi + g'\,F$ and $F$ contains terms with several
fields, Q.E.D.

The effective Lagrangian of \eqn{FLag} is one possible way of parametrizing deviations of the 
Higgs couplings to SM particles; if confirmed, these deviations require new physics models
that are the ultraviolet completion of the set of $d = 6$ operators. However, there are specific
assumptions in considering \eqn{FLag}, namely decoupling of heavy degrees of freedom is assumed
and absence of mass-mixing of the new heavy scalars with the SM Higgs doublet.

We postpone a more detailed discussion of non-decoupling effects to \refS{acdec}; here we note that
\eqn{FLag} comprises all heavy physics effects at scales below $\Lambda$, and in a decoupling
scenario $\Lambda$ is the mass of the additional, heavy, degree of freedom. A typical
non-decoupling scenario is given by the inclusion of a scalar triplet; here higher dimensional
operators are not suppressed by inverse powers of the triplet mass. It is considerably more
difficult to construct a perfectly sensible low-energy effective theory in the non-decoupling 
scenario and the construction is model dependent, \eg it has been shown in
\Bref{SekharChivukula:2007gi} that (in the heavy triplet case) $\Lambda$ is related to the ratio 
of the renormalized triplet VEV to the renormalized doublet VEV. Therefore, additional work
is needed in handling models showing a non-decoupling behavior, \eg looking for the
presence of alternative large parameters. 
%--
\subsection{From the Lagrangian to the $S\,$-matrix\label{FL}}
%--
There are several technical points that deserve a careful comment when we construct $S\,$-matrix 
elements from the Lagrangian of \eqn{FLag}. 
%--
\begin{itemize}
\item{\underline{\bf{Field-scaling, parameter re-definition}}}
\end{itemize}

\begin{figure}[h]
\vspace{2.cm}
\begin{picture}(0,0)(-50,0)
 \SetScale{0.5}
 \SetWidth{1.8}
%--
\DashLine(0,0)(70,0){3}
\Line(75,0)(125,50)
\Line(75,0)(125,-50)
\GCirc(75,0){10}{0.6}
\GCirc(125,50){5}{0}
\GCirc(125,-50){5}{0}
%--
\DashLine(150,0)(220,0){3}
\Line(225,0)(275,50)
\DashLine(225,0)(275,-50){3}
\GCirc(225,0){10}{0.6}
\GCirc(275,50){5}{0}
%--
\DashLine(300,0)(370,0){3}
\DashLine(375,0)(425,50){3}
\Line(375,0)(425,-50)
\GCirc(375,0){10}{0.6}
\GCirc(425,-50){5}{0}
%--
\DashLine(450,0)(520,0){3}
\DashLine(525,0)(575,50){3}
\DashLine(525,0)(575,-50){3}
\GCirc(525,0){10}{0.6}
%--
\Text(0,10)[]{\large $\HB$}
\Text(45,20)[]{\large $\PZB$}
\Text(110,-20)[]{\large $\POpz$}
\Text(135,-40)[]{\large ${\overline\xi}_{\PZB}\,\frac{\OM}{\cth}$}
\Text(210,40)[]{\large ${\overline\xi}_{\PZB}\,\frac{\OM}{\cth}$}
\Text(295,40)[]{\large ${\overline\xi}_{\PZB}\,\frac{\OM}{\cth}$}
\Text(295,-40)[]{\large ${\overline\xi}_{\PZB}\,\frac{\OM}{\cth}$}
\Text(350,2)[]{\large $= 0$}
\Text(65,0)[]{\large $+$}
\Text(140,0)[]{\large $+$}
\Text(210,0)[]{\large $+$}
%--
\end{picture}
\vspace{1.5cm}
\caption[]{Example of Ward-Salvnov-Taylor identity; the grey circle denotes insertion
of $d = 6$ operators, black circles denote the replacement of the polarization vector by $i$ times
the momentum flowing inwards. $\PZB$ and $\POpz$ lines represent Dyson resummed propagators.}
\label{WSTI}
\end{figure}
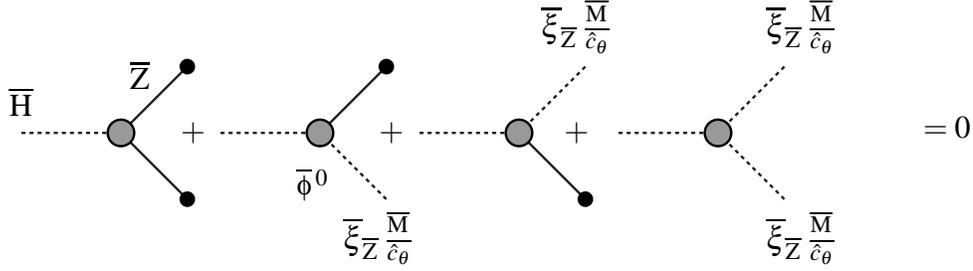

We perform field re-definitions so that all kinetic and mass terms in the Lagrangian of \eqn{FLag}
have the canonical normalization. First we define
%--
\bq
\PH = \HB\,\Bigl[ 1 + \frac{M^2}{\HSs}\,\lpar
a^1_{\PK} + a^3_{\PK} + 2\,a_{\dPK}\rpar\Bigr],
\eq
%--
\bq
\Ppz = \POpz\,\Bigl[ 1 + \frac{M^2}{\HSs}\,\lpar a^1_{\PK} + a^3_{\PK}\rpar\Bigr],
\qquad
\upphi^{\pm} = {\mybar \upphi}^{\pm}\,\Bigl[ 1 + \frac{M^2}{\HSs}\,a^1_{\PK}\Bigr],
\eq
%--
then we introduce new parameters,
%--
\bq
M = \OM\,\lpar 1 + \frac{M^2}{\HSs}\,a^1_{\PK}\rpar,
\qquad
\ctw = {\mybar{c}}_{\theta}\,\lpar 1 - \frac{M^2}{\HSs}\,a^3_{\PK}\rpar.
\eq
%--
Finally we rescale again the fields
%--
\bq
\PZmu = \PZB_{\mu}\,\lpar 1 - 4\,\ACf \,
{\mybar{s}}_{\theta}\,{\mybar{c}}_{\theta}\,a^3_{\PV}\rpar
\qquad
\PA_{\mu} = \myPAB_{\mu}\,\lpar 1 + 4\,\ACf \,
{\mybar{s}}_{\theta}\,{\mybar{c}}_{\theta}\,a^3_{\PV}\rpar,
\eq
%--
redefine the weak-mixing angle as
%--
\bq
{\mybar{c}}_{\theta} = {\hat c}_{\theta}\,\lpar 1 - 4\,\ACf \,
\sth\,\cth\,a^3_{\PV}\rpar,
\eq
%--
and introduce Higgs parameters
%--
\bq
\mhs = \Bigl[ 1 + \ACf \,
\lpar a^1_{\PK} + a^3_{\PK} +2\,a_{\dPK}\rpar\Bigr]\,\OMHs
- 16\,\frac{\OM^4}{g\,\HSs}\,a_{\PK},
\eq
%--
\bq
\beta_{\PH} = \Bigl[ 1 - \ACf \,
\lpar 2\,a^1_{\PK} + a^3_{\PK} + 2\,a_{\dPK}\rpar\Bigr]\,{\mybar{\beta}}_{\PH}
- 4\,\frac{\OM^4}{g\,\HSs}\,a_{\PK}.
\eq
%--
It is worth noting that a different definition of $\Ope^i_{\Pf}$, \ie
%--
\bq
\Ope^1_{\Pf} = g^3\,\PKdag\,\PK\,\POpsiL\,\PK\,\PQtR + \mbox{h. c.},
\qquad
\Ope^2_{\Pf} = g^3\,\PKdag\,\PK\,\POpsiL\,\PK^c\,\PQbR + \mbox{h. c.},
\label{redef}
\eq
%--
requires a re-definition of the $\PQt - \PQb$ bare masses,
%--
\bq
{\mybar{M}}_{\PQt} = \mt - 2\,\srt\,\frac{\OM^2}{\Lambda^2}\,a^1_{\Pf},
\qquad
{\mybar{M}}_{\PQb} = \mb - 2\,\srt\,\frac{\OM^2}{\Lambda^2}\,a^2_{\Pf}.
\label{redemf}
\eq
%--
In the option of \eqn{d6list} the $\PH\Pf\Pf$-Yukawa couplings are
%--
\bqa
\Lag_{\PH\Pf\Pf} &=& \Lag^{\mySM}_{\PH\Pf\Pf}
- \frac{1}{2}\,\frac{\OMs}{\Lambda^2}\,
\Bigl[ g\,\frac{\mt}{\OM}\,\lpar a^3_{\PK}  + 2\,a_{\dPK}\rpar
- 4\,\srt\,a^1_{\Pf}\Bigr]\,\mybar{\PH}\,\PAQt\,\PQt
\nl
{}&-& \frac{1}{2}\,\frac{\OMs}{\Lambda^2}\,
\Bigl[ g\,\frac{\mb}{\OM}\,\lpar a^3_{\PK}  + 2\,a_{\dPK}\rpar
- 4\,\srt\,a^2_{\Pf}\Bigr]\,\mybar{\PH}\,\PAQb\,\PQb
\eqa
%--
while, following \eqn{redef} and \eqn{redemf}, we obtain
%--
\bq
\Lag_{\PH\Pf\Pf} = \Lag^{\mySM}_{\PH\Pf\Pf}
- \frac{1}{2}\,g\,\frac{\mybar{M}_{\PQt} \OM}{\Lambda^2}\,\lpar a^3_{\PK}  + 2\,a_{\dPK}\rpar\,
\mybar{\PH}\,\PAQt\,\PQt 
- \frac{1}{2}\,g\,\frac{\mybar{M}_{\PQb} \OM}{\Lambda^2}\,\lpar a^3_{\PK}  + 2\,a_{\dPK}\rpar\,
\mybar{\PH}\,\PAQb\,\PQb. 
\eq
%--
\begin{itemize}
\item{\underline{\bf{gauge-fixing term}}}
\end{itemize}
%--
We define a modified gauge-fixing term for the $\PW,\PZ\,$-fields,
%--
\bq
C^{\pm} = -\,\pdmu\,\PWpmmu + \xiW\,M\,\upphi^{\pm},
\qquad
C_{\PZ} = -\,\pdmu\,\PZmu + \xiZ\,\frac{M}{\ctw}\,\Ppz
\eq
%--
where the gauge-parameters are
%--
\bq
\xiW = 1 - 2\,\ACf \,a^1_{\PK},
\qquad
\xiZ = 1 - 2\,\ACf \,\lpar a^1_{\PK} + a^3_{\PK}\rpar.
\eq
%--
It is straightforward to show that
%--
\bq
C^{\pm} = -\,\pdmu\,\PWpmmu + \,\OM\,\POppm,
\qquad
C_{\PZ} = -\,\frac{1}{\OxiZ}\,\pdmu\,\PZB_{\mu} 
+ \OxiZ\,\frac{\OM}{{\hat c}_{\theta}}\,\POpz,
\eq
%--
where the gauge parameter is
%--
\bq
\OxiZ = 1 + 4\,\ACf \,
{\mybar{s}}_{\theta}\,{\mybar{c}}_{\theta}\,a^3_{\PV}.
\label{gfpar}
\eq
%--
Note taht the photon gauge-fixing term remains unchanged, \ie
%--
\bq
C_{\PA} = -\,\pdmu\,\myPAB_{\mu}.
\eq
%--
With our choice for the scaling factors, the parameter redefinition and the form of the 
gauge-fixing term it follows that the part of the Lagrangian which is quadratic in the (bosonic) 
fields reads:
%--
\bqa
\Lag^{\bos}_2 &=&
       - \pdmu \PWpnu \, \pdmu \PWmnu 
       - \OMs \, \PWpmu \, \PWmmu 
       - \frac{1}{2} \, \pdmu\PZB_{\nu} \, \pdmu\PZB_{\nu} 
       - \frac{1}{2}\,\frac{\OMs}{\cths} \, \PZB_{\mu} \, \PZB_{\mu}           
       - \frac{1}{2} \, \pdmu\myPAB_{\nu} \, \pdmu\myPAB_{\nu} 
\nl
{}&    - &\frac{1}{2} \, \pdmu \HB \pdmu \HB
       - \frac{1}{2}\,\OMHs \, \HB^2
       - \pdmu\Ppp \, \pdmu\Ppm
       - \OMs \, \Ppp \, \Ppm
       - \frac{1}{2} \, \pdmu\POpz \, \pdmu\POpz
       - \frac{1}{2} \, \OxiZs \, \frac{\OMs}{\cths} \, \lpar \POpz\rpar^2     
\nl
{}&    + &4\,\ACf \,\cth\,\sth\,\Bigl[ 
         a^3_{\PV} \,\lpar 
         \pdmu \PZB_{\nu}\,\pdnu \PZB_{\mu} 
       - \pdmu \myPAB_{\nu}\,\pdnu \myPAB_{\mu}\rpar 
       - 2\,\ep^{\mu\nu\alpha\beta}\,a^3_{\essV} \lpar 
         \pdmu \PZB_{\nu}\,\partial_{\alpha} \PZB_{\beta} 
       - \pdmu \myPAB_{\nu}\,\partial_{\alpha} \myPAB_{\beta} \rpar\Bigr]
\nl
{}&    + &4\,(1 - 2\,\sth^2)\,\ACf \,\Bigl[ 
         a^3_{\PV} \lpar \pdmu \PZB_{\nu}\,\pdmu \myPAB_{\nu} 
       - \pdmu \PZB_{\nu}\,\pdnu \myPAB_{\mu} \rpar 
       + 2\,\ep^{\mu\nu\alpha\beta}\,a^3_{\essV} \, 
         \pdmu \PZB_{\nu}\,\partial_{\alpha} \myPAB_{\beta}\Bigr] 
\eqa
%--
Therefore, {\bf{kinetic and mass terms are SM-like}}, and the bare $\uprho$ parameter is 
$\ord{1/\Lambda^2}$, by construction.
%--
\begin{itemize}
\item{\underline{\bf{Dyson resummed propagators}}}
\end{itemize}
%--
Dyson resummed propagators are crucial for discussing several issues, from renormalization
to Ward-Slavnov-Taylor (WST) identities~\cite{Veltman:1970nh,Taylor:1971ff,Slavnov:1972fg}. 
Consider the $\PW$ or $\PZ$ self-energy; in the SM we have
%--
\bq
\Pi^{\PV\PV}_{\mu\nu}(p^2) = \Pi^{\PV\PV}_0(p^2)\,\delta_{\mu\nu} +
\Pi^{\PV\PV}_1(p^2)\,p_{\mu} p_{\nu}.
\eq
%--
Once $d = 6$ operators are added the $\PW$ Dyson resummed propagator remains unchanged, \ie
%--
\bq
\oD^{\PWW}_{\mu\nu} =
\frac{\delta_{\mu\nu}}{p^2 + \OMs - \Pi^{\PWW}_0} +
\frac{\Pi^{\PWW}_1\,p_{\mu} p_{\nu}} 
{\lpar p^2 + \OMs - \Pi^{\PWW}_0\rpar\,
 \lpar p^2 + \OMs - \Pi^{\PWW}_0 - \Pi^{\PWW}_1\,p^2\rpar},
\eq
%--
while the $\PZB$ propagator changes as follows:
%--
\bq
\Pi^{\PZB\PZB}_1 \qquad \to \qquad \Pi^{\PZZ}_1 + 4\,\ACf \,\sth\,\cth\,a^3_{\PV}.
\eq
%--
For the $\phib$ propagators we get
%--
\bq
\oD^{\;\POpz\POpz}(p^2)= 
\frac{1}{p^2 + {\overline\xi}^2_{\PZ}\,\frac{\OMs}{\cths}},
\qquad
\oD^{\;\Ppp\Ppm}(p^2)= \frac{1}{p^2 + \OMs}.
\eq
%--
with the gauge parameter defined in \eqn{gfpar}.
%--
\begin{itemize}
\item{\underline{\bf{WST identities}}}
\end{itemize}
%--
With the Feynman rules developed above we can prove WST identities; we show
an example in \refF{WSTI} where one should take into account that all lines must be on-shell
otherwise there are additional terms involving FP-ghost lines, \ie BRST-invariance requires
also effective operators involving ghost-fields.
%--
%--
\begin{itemize}
\item{\underline{\bf{Wave-function factors}}}
\end{itemize}
%--
Due to the rescaling of the fields each external leg in a $S\,$-matrix element has to
be multiplied by a factor; the argument is general, given a Lagrangian
%--
\bq
\Lag = Z^{-2}\,\upphi\,\Delta^{-1}\,\upphi + J\,\upphi
\eq
%--
we have to normalize the source $J$ in such a way that the residue of the two-point $S\,$-matrix
element is one; therefore we fix $J \to Z^{-1}\,J$ for the $S\,$-matrix element containing one
external $\upphi\,$-line. We define $Z_i= 1 + \delta Z_i$ and obtain
%--
\bq
\delta Z_{\PH} = \ACf \,\lpar a^1_{\PK} + a^3_{\PK} + 2\,a_{\dPK}\rpar,
\quad
\delta Z_{\Ppz} = \ACf \,\lpar a^1_{\PK} + a^3_{\PK} \rpar,
\quad
\delta Z_{\upphi} = \ACf \,a^1_{\PK},
\label{WFF1}
\eq
%--
\bq
\delta Z_{\PW} = 0, \quad
\delta Z_{\PZ} = -4\,\ACf \,\frac{\sth}{\cth}\,a^3_{\PV},
\quad
\delta Z_{\PA} = 4\,\ACf \,\frac{\sth}{\cth}\,a^3_{\PV}.
\label{WFF2}
\eq
%--
\subsection{Nature of $d= 6$ operators \label{natop}}
%--
The $d= 6$ operators are supposed to arise from a local Lagrangian, containing heavy
degrees of freedom, once the latter are integrated out. Of course, the correspondence
Lagrangians $\,\to\,$ effective operators is not bijective since many different
Lagrangians can give raise to the same operator. Nevertheless these operators are of
two different origins~\cite{Arzt:1994gp}:
%--
\begin{itemize}
\item $T\,$-operators are those that arise from the tree-level exchange of some heavy degree
of freedom
\item $L\,$-operators are those that arise from loops of heavy degrees of freedom.
\end{itemize} 
%--
The $L\,$-operators are usually not included in the analysis. The accuracy at which results 
should be presented is given by
%--
\bq
\Amp = \Amp^{\myLO}_{\mySM} + \Amp^{\myNLO}_{\mySM} + \Amp^{\myLO}_{d=6},
\eq
%--
where LO means the first order in perturbation theory where the amplitude receive a contribution.
To be precise, if the underlying theory is weakly-coupled operators containing field-strength 
tensors cannot be $T\,$-operators, and their Wilson coefficients are $1/(16\,\pi^2)$
suppressed. If only $T\,$-operators (coefficients of $\ord{1}$) are included only
$14$ out of $34$ entries in Table~2 of \Bref{Grzadkowski:2010es} are relevant.
There is another caveat: $d = 6$, $L\,$-operators have Wilson coefficients 
$\sim\,1/(16\,\pi^2)$ and $d = 8$, $T\,$-operators are $\sim\,v^2/\Lambda^2$; therefore,
below $\approx 3\UTeV$ one should include both of them or none of them.
%--
\subsubsection{Insertion of $d = 6$ operators in loops}
%--
The question remains on insertion of $d= 6$ operators in SM loop diagrams. This is better 
discussed in terms of a concrete example: consider the Lagrangian~\cite{Bonnet:2011yx}
%--
\bq
\Lag = \Lag_{\mySM} - \frac{1}{2}\,\pdmu S\,\pdmu S
- \frac{1}{2}\,\mPs\,S^2 + \mu_{\PS}\,\PKdag\,\PK\,S,
\label{NPL}
\eq
%--
where $S$ is a heavy (scalar) singlet The interaction is
%--
\bq
\Lag_{\Lint} = \frac{1}{2}\,\mu_{\PS}\,\lpar \PH^2 + \Ppz \Ppz + 2\,\Ppp \Ppm\rpar\,S.
\eq
%--
In the limit $M_{\PS} \to \infty$ we have
%--
\bq
\Lag \to \Lag^{\myLO}_{\mySM} + \frac{\mu^2_{\PS}}{\mPs}\,\PKdag \PK +
\frac{\mu^2_{\PS}}{M^4_{\PS}}\,\Ope_{\dPK}.
\label{Lint}
\eq
%--
The $d= 4$ operator in \eqn{Lint} can be absorbed through a parameter redefinition, and we are 
left with a contribution to the $d= 6$ operator $\Ope_{\dPK}$. Clearly, $\Ope_{\dPK}$ 
(as well as $\Ope_{\PK}$, $\Ope^1_{\PK}$ and $\Ope^3_{\PK}$) is a 
$T\,$-operator~\cite{Bonnet:2011yx,Arzt:1993gz,Arzt:1994gp}.
%--
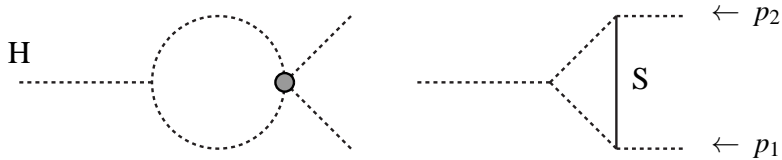
\begin{figure}[h]
\vspace{2.cm}
\begin{picture}(0,0)(-70,0)
 \SetScale{0.5}
 \SetWidth{1.8}
%--
\DashLine(0,0)(100,0){3}
\DashCArc(150,0)(50,0,360){3}
\DashLine(200,0)(250,50){3}
\DashLine(200,0)(250,-50){3}
\GCirc(200,0){7}{0.6}
%--
\DashLine(300,0)(400,0){3}
\DashLine(400,0)(450,50){3}
\DashLine(400,0)(450,-50){3}
\Line(450,50)(450,-50)
\DashLine(450,50)(500,50){3}
\DashLine(450,-50)(500,-50){3}
%--
\Text(0,10)[]{\large $\PH$}
\Text(235,0)[]{\large $\PS$}
\Text(275,25)[]{$\leftarrow\;p_2$}
\Text(275,-25)[]{$\leftarrow\;p_1$}
%--
\end{picture}
\vspace{1.cm}
\caption[]{The three-point function $\PH^3$ with the insertion of the ${\cal O}_{\dPK}$
operator (left) and the same contribution in the full Lagrangian of \eqn{NPL}.}
\label{HC}
\end{figure}
%--

Imagine we want to compute the $\PH^3$ Green function: we analyze the ultraviolet (UV) behavior 
of the two diagrams in \refF{HC}. In the effective theory (left diagram) there is an UV 
divergence and one option would be to subtract it by introducing counterterms in $\Lag_{d=6}$. 
However, this shows how the insertion of local operators of higher dimensionality in SM diagrams 
is not really consistent since, in the full theory, the corresponding diagram is not divergent. 
If we introduce $\HSs = \mu^2_{\PS}/m^4_{\PS}$ the diagram behaves like 
$\Lambda^{-2}\,\ln \Lambda$, \ie the divergence is controlled by the heavy mass.
From this point of view it is important to stress that one should avoid using a cutoff
procedure with the dimensionful parameter $\Lambda$.
Computing with dimensional regularization gives a different pole structure reflecting the
different counterterms in the full and effective theory. This difference is independent of
infrared (IR) physics, since both theories have the same IR behavior.
As we have seen, there can also be logarithmic dependence on $\Lambda$; if these logarithms are 
included they must be summed. 

To give an example we consider again the two diagrams of \refF{HC} with $s = - (p_1 + p_2)^2$. 
In the effective theory the insertion of $\Ope_{\dPK}$ (left diagram in \refF{HC}) produces
%--
\bq
I_{\eff} = \frac{3}{4}\,g\,\frac{\OMHs}{\OM \Lambda^2}\,
\mu_{\ssR}^{\ep}\,\int d^n q\,\frac{\lpar q + p_1\rpar^2}
               {\lpar q^2 + \OMHs\rpar\,\lpar\lpar q + p_1 + p_2\rpar^2 + \OMHs\rpar},
\eq
%--  
where $n= 4 - \ep$. Suppose that we use a cut-off regularization, the integral is $\ord{1}$ for 
$\Lambda \to \infty$ but the same is true for all integrals containing the insertion
of $\Ope^n_{\dPK}$ operators; therefore, all these diagrams are of the same order and cannot
be neglected. In dimensional regularization (DR) we obtain
%--
\bq
I^{\DR}_{\eff} = \frac{3}{4}\,g\,\frac{\OMHs}{\OM \Lambda^2}\,\Bigl[
\lpar \frac{1}{2}\,s - 3\,\OMHs\rpar\,\lpar \frac{1}{\mybar{\ep}} - \ln \frac{\mu^2_{\ssR}}{s}\rpar
+ \quad \mbox{finite part} \;\Bigr].
\eq
%--
where $1/{\mybar{\ep}} = 2/(4-n) - \gamma - \ln \pi$ and $\mu_{\ssR}$ is the
renormalization scale. In principle, we could add counterterms (in the $\mybar{MS}$ scheme)
to remove the UV pole and make a choice for the scale, $\mu_{\ssR}$, which minimizes the 
remaining logarithms in the UV finite part. After subtracting the UV pole we can say that the
insertion of a $d = 6$ operator produces a result
%--
\bq
I^{\ren}_{d = 6} \sim \frac{\OMHs}{\HSs}\,\ln\mu_{\ssR}.
\eq
%--
The insertion of a $d = 8$ operator, always working in dimensional regularization, gives
%--
\bq
I^{\ren}_{d = 8} \sim \frac{\OMHq}{\Lambda^4}\,\ln\mu_{\ssR},
\eq
%--
\etc Note that, with cutoff regularization, both integrals would be of ${\cal O}(1)$.  
Note that in a mass-independent scheme like $\overline{\rm MS}$ the conditions for the
decoupling theorem are not satisfied. Furthermore, the logarithms of the renornalization mass
may become large. In principle, the problem can be solved but the solution requires matching
conditions (for a discussion see \Bref{Manohar:1996cq}). 

With the full theory at our disposal we compute
%--
\bq
I_{\full} = -\,\frac{3}{2}\,g\,\frac{\OMHs \mu^2_{\PS}}{\OM}\,
\int d^n q\,\frac{1}
               {\lpar q^2 + \OMHs\rpar\,\lpar\lpar q + p_1\rpar^2 + \mPs\rpar\,
                \lpar\lpar q + p_1 + p_2\rpar^2 + \OMHs\rpar}.
\label{startP}
\eq
%--
Working (for simplicity) with $\OMHs \muchless s \muchless \mPs$ we obtain
%--
\bq
I_{\full} = \,\frac{3}{2}\,g\,\frac{\OMHs \mu^2_{\PS}}{\OM s}\,\Bigl[
\zeta(2) - \li{2}{1 + \frac{s + i\,0}{\mPs}} \Bigr].
\eq
%--
We can identify $\Lambda = \mPs/\mu_{\PS}$, expand in $s/\mPs$, and obtain
%--
\bq
I_{\full} = -\,\frac{3}{2}\,g\,\frac{\OMHs \mu^2_{\PS}}{\OM \mPs}\,[
1 - \frac{1}{4}\,\frac{s}{\mPs} + \lpar 1 - \frac{1}{2}\,\frac{s}{\mPs} \rpar\,
\ln \frac{-s - i\,0}{\mPs} + {\cal O}\lpar \frac{s^2}{M^4_{\PS}}\rpar \Bigr].
\eq
%-- 
The first term in $I_{\full}$ reproduces the $d = 4$ operator of \eqn{Lint} while the
second term corresponds to the $d = 6$, $\Ope_{\dPK}$ operator. There is no UV divergence 
in $I_{\full}$ and the logarithm is uniquely fixed.

An alternative way to understand the two different approaches is the following: we start
from \eqn{startP} and expand in the integrand
%--
\bq
\frac{1}{\lpar q + p_1\rpar^2 + \mPs} = \frac{1}{\mPs}\,\Bigl[
1 - \frac{\lpar q + p_1\rpar^2}{\mPs} + \,\cdots\,\Bigr],
\eq
%--
which is equivalent to inserting $d \ge 4$ operators or introduce Feynman parameters:
%--
\bqa
J &=& \int d^n q\,\frac{1}
      {q^2\,\lpar\lpar q + p_1\rpar^2 + \mPs\rpar\,\lpar q + p_1 + p_2\rpar^2}
\nl
{}&=&
\intfxy{x}{y}\,\Bigl[ \mPs\,\lpar x - y\rpar - s\,y\,\lpar 1 - x\rpar\Bigr]^{-1}
\eqa
%-- 
use a Mellin-Barnes representation, and expand as follows ($M_{\PS} \to \infty$):
%--
\bqa
J &=& \frac{1}{2 \pi i}\,\int_{-\,\infty}^{+\,\infty}\,dv
\,\lpar \mPs\rpar^{v-1}\,\lpar - s\rpar^{-v}\,\intfxy{x}{y}\,B\lpar v\,,\,1 - v\rpar\,
\lpar x - y\rpar^{v-1}\,y^{-v}\,\lpar 1 - x\rpar^{-v}
\nl
{}&=& \frac{1}{2 \pi i}\,\frac{1}{\mPs}\,\int_{-\,\infty}^{+\,\infty}\,dv\,
\frac{\Gamma^2(s)\,\Gamma^2(1-s)}{1-s}\,\lpar \frac{-\,\mPs}{s}\rpar^v.
\eqa
%--
Here $B(x,y)$ is the Euler beta-function. Using the well know Laurent and Taylor expansions
of the Euler gamma-function we obtain the result summing over the poles at $s = - n$:
%--
\bq
J = \sum_{n=0}^{\infty}\,\frac{1}{n+1}\,\frac{\lpar -\,s\rpar^n}{\lpar \mPs\rpar^{n+1}}\,
\Bigl[ \frac{1}{n+1} + \ln \lpar -\,\frac{\mPs}{s}\rpar \Bigr].
\eq
%--
The result is manifestly UV finite, term-by-term, and has the correct structure of logarithms.
%--
\subsubsection{Admissible operators}
%--
Missing a candidate for the BSM Lagrangian, we will not deal with renormalization of composite 
operators; therefore, we will not include local operators in loops. To be more precise we will use
the following set of rules:
%--
\begin{enumerate}

\item operators altering the UV power-counting of a SM diagram are non-admissible

\item operators that do not change the UV power-counting are admissible only
in a very specific case: we say that a set of SM diagrams is UV-scalable w.r.t. a 
combination of $d= 6$ operators if

\begin{itemize}

\item their sum is UV finite
\item all diagrams in the set are scaled by the same combination of $d= 6$ operators.

\end{itemize}

\end{enumerate}
%--
To explain with one specific example, let us consider the $\PH\PWW$ vertex
with off-shell lines and no wave-function factor inserted:
%--
\bqa
V_{\PH\PWW}^{\mu\nu} &=&
     - g\,\OM\,
        \Bigl[ 1 + \lpar a^3_{\PK} - 2\,a^1_{\PK} + 2\,a_{\dPK} \rpar\,\ACf  
        \Bigr]\,\delta^{\mu\nu}
      + a^4_{\PK}\,\frac{\OM}{\HSs}\,P.P\,\delta^{\mu\nu}
\nl
{}&+& 8\,a^1_{\PV}\,\frac{\OM}{\HSs}\,T^{\mu\nu} 
      - a^4_{\PK}\, \frac{\OM}{\HSs}\,
        \lpar p^{\mu}_1\,p^{\nu}_1 + 2\,p^{\nu}_1\,p^{\mu}_2 + p^{\mu}_2\,p^{\nu}_2 \rpar
      + 16\,a^1_{\essV}\,\frac{\OM}{\HSs}\, \ep^{\alpha\beta\mu\nu}\,p_{1\alpha}\,p_{2\beta} 
\eqa
%--
with $T^{\mu\nu} = p^{\mu}_2\,p^{\nu}_1 - \spro{p_1}{p_2}\,\delta^{\mu\nu}$ and
$P= p_1 + p_2$.
Consider the one-loop diagram contributing to $\PH \to \PGg \PGg$ containing a $\PW$
loop: the operators $\Ope^4_{\PK}$ and $\Ope^1_{\PV}, \Ope^1_{\essV}$ 
change the UV power-counting of the original SM diagram and are non-admissible.
           
\begin{figure}[ht]
\vspace{2.cm}
\begin{picture}(0,0)(-70,0)
 \SetScale{0.5}
 \SetWidth{1.8}
%--
\DashLine(0,0)(50,0){3}
\Line(50,0)(100,50)
\Line(50,0)(100,-50)
\Line(100,50)(100,-50)
\Photon(100,50)(150,50){4}{4}
\Photon(100,-50)(150,-50){4}{4}
\Text(0,10)[]{\large $\PH$}
\Text(76,-12)[]{\large $\PW/\upphi/\PXpm$}
%--
\DashLine(200,0)(250,0){3}
\CArc(275,0)(25.,0.,360)
\Photon(300,0)(350,50){4}{4}
\Photon(300,0)(350,-50){4}{4}
\Text(100,10)[]{\large $\PH$}
\Text(135,-30)[]{\large $\PW/\upphi$}
%--
\DashLine(400,0)(450,0){3}
\CArc(475,25)(28.,0.,360)
\Photon(490,50)(550,50){4}{4}
\Photon(450,0)(550,-50){4}{4}
\Text(200,10)[]{\large $\PH$}
\Text(270,5)[]{\large $\PW/\upphi$}
%--
\end{picture}
\vspace{1.cm}
\caption[]{The three families of diagrams contributing to the bosonic amplitude for
$\PH \to \PGg\PGg$; $\PW/\upphi$ denotes a $\PW\,$-line or a $\upphi\,$-line. $\PXpm$ denotes
a FP-ghost line}
\label{top}
\end{figure}
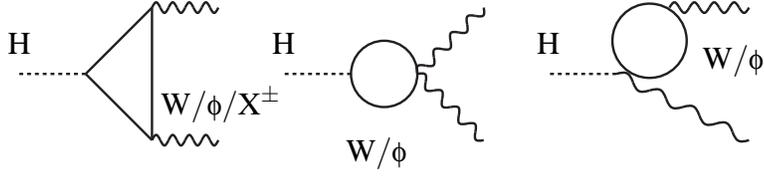

In the one-loop (bosonic) amplitude for $\PH \to \PGg \PGg$ there are three different contribution,
a $\PW\,$-loop, a charged $\upphi\,$-loop and a mixed $\PW-\upphi$ loop. We find
%--
\bqa
V_{\PH\PW\upphi}^{\nu} &=&
         i\,g\,\Bigl[ 1 + \lpar a^3_{\PK} - 2\,a^1_{\PK} + 2\,a_{\dPK}\rpar \,\ACf  
         \Bigr]\,p^{\nu}_1 
       + i\,a^4_{\PK}\, \frac{\spro{p_2}{p_2}}{\HSs}\,p^{\nu}_1 
\nl
{}&+& \frac{i}{2}\,g\,
         \Bigl[ 1 + \lpar a^3_{\PK} + 2\,a_{\dPK} \rpar\,\ACf  \Bigr]\,p^{\nu}_2
       - i\,a^4_{\PK}\, \frac{\spro{p_1}{p_2}}{\HSs}\,p^{\nu}_2
\nl
V_{\PH\upphi\upphi} &=&
       - \frac{1}{2}\,g\, \frac{\OMHs}{\OM}\,
         \Bigl[ 1 + \lpar a^3_{\PK} - 2\,a^1_{\PK} + 2\,a_{\dPK} \rpar\,\ACf \Bigr]
       + g\,\OM\,a^1_{\PK}\, \frac{\spro{p_1}{p_1} + \spro{p_2}{p_2}}{\HSs}
\nl
{}&+& g\,\OM\, \lpar a^3_{\PK} - a^1_{\PK} + 2\,a_{\dPK} \rpar \, 
         \frac{\spro{P}{P} + \OMHs}{\HSs}
\eqa
%--
It is straightforward to conclude that the SM one-loop, bosonic, amplitude for
$\PH \to \PGg \PGg$ with on-shell Higgs line is UV-scalable w.r.t. the combination
%--
\bq
C_{\bos}= \ACf \,\lpar a^3_{\PK} - 2\,a^1_{\PK} + 2\,a_{\dPK}\rpar,
\eq
%--
which could be admissible. However, in the one-loop amplitude we also have FP-ghost loops with
vertices (see \eqn{FPLag})
%--
\bq
V_{\PH\PAXpm\PXpm} = 
 - \frac{1}{2}\,g\,\OM\,\Bigl[ 1 + \lpar a^3_{\PK} + 2\,a_{\dPK}\rpar\,\ACf \Bigr].
\eq
%--
Therefore the bosonic component is only UV-scalable w.r.t. the combination
%--
\bq
C^1_{\bos}= \ACf \,\lpar a^3_{\PK} + 2\,a_{\dPK}\rpar.
\eq
%--
Similarly, we consider the $\PGg\PWW$, $\PGg\PW\upphi$, $\PGg\upphi\upphi$ and
$\PGg{\mybar{\PX}}^{\pm}\PX^{\pm}$ vertices, which also appear in the one-loop bosonic amplitude 
for $\PH \to \PGg\PGg$, and conclude that the latter is UV-scalable w.r.t. the combination
%--
\bq
C^2_{\bos} = \ACf \,\frac{\cth}{\sths}\,
\lpar 4\,\sth\,a^3_{\PV} + \cth\,a^3_{\PK}\rpar,
\eq
%--
which is also admissible. Obviously, the wave-function factors of \eqns{WFF1}{WFF2}
are also admissible. 
To be more precise, the one-loop bosonic amplitude for $\PH \to \PGg\PGg$ is made
of three different families of diagrams, shown in \refF{top}.
We find that the $\PGg\PGg\PW\PW$, $\PGg\PGg\PW\upphi$ and $\PGg\PGg\upphi\upphi$ vertices
are all UV-scalable w.r.t. $2\,C^2_{\bos}$. Furthermore, the vertex $\PGg\PH\PW\upphi$
is UV-scalable w.r.t. $C^1_{\bos} + C^2_{\bos}$. The underlying algebra is such that
the quadrilinear vertex with two $\PGg$s is equivalent to the square of the trilinear vertex
with one $\PGg$ (to $\ord{1/\Lambda^2}$) and the quadrilinear vertex with one $\PH$ 
is equivalent (to the same order) to the product of the two trilinear vertices, with
a $\PGg$ and with a $\PH$. As a consequence, there is a non-trivial scaling factor which is
admissible, not spoiling the UV behavior.
 
%--
%--

\begin{figure}[hb]
\vspace{2.cm}
\begin{picture}(0,0)(-50,0)
 \SetScale{0.5}
 \SetWidth{1.8}
%--
\DashLine(0,0)(100,0){3}
\Line(100,0)(150,50)
\Line(100,0)(150,-50)
\Line(150,50)(150,-50)
\Photon(150,50)(250,50){4}{4}
\Photon(150,-50)(250,-50){4}{4}
%--
\DashLine(400,0)(500,0){3}
\Photon(500,0)(600,50){4}{4}
\Photon(500,0)(600,-50){4}{4}
\GCirc(500,0){7}{0.6}
%--
\Text(170,0)[]{\Large $\Longrightarrow$}
%--
\end{picture}
\vspace{1.cm}
\caption[]{Example of diagram giving a contribution to the $d= 6$ operator of type $L$.
Solid lines represent colored scalar fields, \eg transforming in the 
$\lpar 8\,,\,2\,,\,\frac{1}{2}\rpar$ representation of $SU(3)\,\otimes\,SU(2)\,\otimes\,U(1)$.}
\label{AG}
\end{figure}
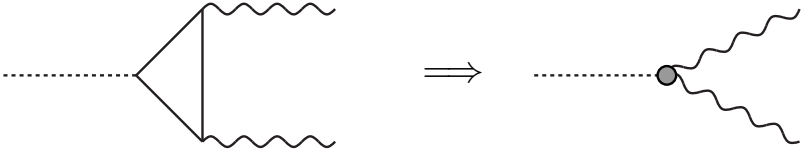

The fermionic amplitude for $\PH \to \PGg \PGg$ contains a top-quark loop and
a bottom-quark loop. The top contribution is UV-scalable w.r.t. the combination
%--
\bq
C^{\PQt}_{\fer} =
  - \frac{1}{2}\,g\,\frac{\mt}{\OM}\,\Bigl[ 1 + \lpar a^3_{\PK} + 2\,a_{\dPK}\rpar\,
\ACf \Bigr] 
  + \frac{1}{4\,\srt}\,\ACf\,\Bigl[ 2\,a^1_{\Pf} 
  + \frac{\spro{P}{P} - 2\,\mts}{\OMs}\,a^3_{\Pf} \Bigr],
\eq
%--
while for the bottom-quark we have
%--
\bq
C^{\PQb}_{\fer} =
  - \frac{1}{2}\,g\,\frac{\mb}{\OM}\,\Bigl[ 1 + \lpar a^3_{\PK} + 2\,a_{\dPK}\rpar\,
\ACf \Bigr] 
  + \frac{1}{4\,\srt}\,\ACf\,\Bigl[ 2\,a^2_{\Pf} 
  + \frac{\spro{P}{P} - 2\,\mbs}{\OMs}\,a^4_{\Pf} \Bigr].
\eq
%--
One example of $L\,$-operator is given by in \refF{AG} with contributions from heavy
colored scalar fields transforming in a $\lpar C\,,\,T\,,\,Y\rpar$ representation of
$SU(3)\,\otimes\,SU(2)\,\otimes\,U(1)$, \eg the $\lpar 8\,,\,2\,,\,1/2\rpar$ 
representation of~\cite{Manohar:2006ga,Manohar:2006gz,Bonciani:2007ex,Kribs:2012kz}.
%--
%--
Since the additional colored scalar (weak-isospin) doublet contains also an electrically 
charged scalar (and two neutral scalars) it will contribute to the decay 
$\PH \to \PGg \PGg$~\cite{Aglietti:2006tp}. As long as the scalars are in a representation 
$\lpar C\,,\,T\,,\,Y\rpar$ such that ${\mybar{C}}\,\otimes\,C \;\ni\; 8$ there will also be 
a contribution to gluon fusion.
%--
\subsection{Effective theory and renormalization \label{acren}}
%--
There are two conceptual frameworks to discuss renormalization and effective theories. In one case
we are only interested in setting up an expansion in power of $E/\Lambda$ where 
$\Lambda$ is the cutoff and $E$ is the scale relevant for a given set of processes. 

Counterterms are introduced to remove UV-divergences and, in presence of $d > 4$ operators, 
an infinite number of them is required. However, once the requested precision of the calculation 
is fixed only a limited number of term is needed. 

This is not the goal for Higgs physics where we want to search for new physics without committing 
to a particular extension of the SM. The effective theory should simply capture the low-energy 
effects of the underlying, BSM, theory and must be replaced by a new one when $E$ is approaching 
$\Lambda$, where it should be discarded.
Having this difference in mind we proceed in discussing renormalization.

The processes $\Pg \Pg \to \PH$ and $\PH \to \PGg \PGg$ are special in the sense that
there is no tree-level coupling and therefore the NLO (one-loop) amplitude is
UV finite. This is not the case for other processes, \ie $\PH \to \PAQb \PQb$ etc.
In general we will have
%--
\bq
{\cal A} = f\lpar \{a_{\AC}\}\rpar\,\Bigl[ A_{\myLO}\lpar \{p_0\}\rpar +
           A_{\myNLO}\lpar \{p_0\}\rpar\Bigr] +
           A_{\AC}\lpar \{a_{\AC}\,,\,p_0\}\rpar,
\eq
%--
where $\{p_0\}$ is the set of bare parameters (masses and couplings), $\{a_{\AC}\}$ a set of 
effective parameters; furthermore $A_{\myLO}(A_{\myNLO})$ is the LO(NLO) SM amplitude.
Since $A_{\NLO}$ contains UV divergences we introduce counterterms
%--
\bq
p_0 = p_{\ren} + \delta Z_p,
\eq
%--
where $p_{\ren}$ is the renormalized parameter and $\delta Z_p$ contains counterterms.
If $A'$ denotes the derivative of the amplitude w.r.t. parameters we obtain
%--
\bq
{\cal A} = f\lpar \{a_{\AC}\}\rpar\,\Bigl[
           A_{\myLO}\lpar \{p_{\ren}\}\rpar +
           A'_{\myLO}\lpar \{p_{\ren}\}\rpar\,\otimes\,\{Z_p\} + 
           A_{\myNLO}\lpar \{p_{\ren}\}\rpar\Bigr] +
           A_{\AC}\lpar \{a_{\AC}\,,\,p_{\ren}\}\rpar.
\eq
%--
The combination
%--
\bq
A'_{\myLO}\lpar \{p_{\ren}\}\rpar\,\otimes\,\{Z_p\} + A_{\myNLO}\lpar \{p_{\ren}\}\rpar 
\eq
%--
is now UV finite. Note that we have replaced $p_0 \to p_{\ren}$ in $A_{\AC}$ because 
in the full theory $A_{\AC}$ is of the {\em same order} of $A_{\myNLO}$, \ie
renormalization of $a_{\AC}$ can only be discussed in the context of the full theory.
In a sense, the $a_{\AC}$ parameters are already the renormalized ones.

There is a final step in the procedure, finite renormalization, where we have to relate
renormalized parameters to physical quantities (\eg $e^2 = g^2\sths = \alpha/(4\,\pi)$),
%--
\bq
p_{\ren} = p_{\esp} + F\lpar \{p_{\esp}\} \rpar. 
\eq
%--
This substitution induces another shift in the amplitude
%--
\bq
A_{\myLO}\lpar \{p_{\ren}\rpar \to 
A_{\myLO}\lpar \{p_{\esp}\rpar + 
A'_{\myLO}\lpar \{p_{\esp}\rpar\,F\lpar \{p_{\esp}\} \rpar,
\eq
%--
with $p_{\ren} = p_{\esp}$ in both $A_{\myNLO}$ and $A_{\AC}$. This set of replacements
completely defines our renormalization procedure.  

A subtle point is the following: in the process $\PH \to \PGg \PGg$ we have a bosonic component of 
$A_{\myLO}$ and a fermionic one and both are UV finite. Therefore, as long as all tree-vertices
in the bosonic part are scaled with the same factor, we would like to have
%--
\bq
{\cal A} = f_{\bos}\lpar \{a_{\AC}\}\rpar\,A^{\bos}_{\myLO}\lpar \{p_0\}\rpar +
           f_{\fer}\lpar \{a_{\AC}\}\rpar\,A^{\fer}_{\myLO}\lpar \{p_0\}\rpar +
           A_{\AC}\lpar \{a_{\AC}\,,\,p_0\}\rpar.
\eq
%--
LO implies one-loop diagrams where the splitting bosonic-fermionic has a meaning. Once we try
to go to NLO (\ie two-loops) the splitting is not definable and renormalization is requested,
\ie one has to insert counterterms in the one-loop diagrams. Clearly, an arbitrary
scaling of the two LO components kills two-loop UV finiteness (at least in the electroweak
sector). The two-loop electroweak corrections to $\PH \to \PGg\PGg$ are $-1.65\%$ at 
$M_{\PH} = 125\UGeV$~\cite{Actis:2008ts}, therefore neglecting them is tolerable but the 
internal inconsistency remains. The effect on $\Pg \Pg \to \PH$ is larger, $\ord{5\%}$.

To conclude this section we compare the BSM scenario with heavy degrees of freedom and the
SM one in the limit of infinitely massless top-quark. In this case we have a coupling 
$\PH\Pg\Pg$ of the form
%--
\bq
\Lag_{\int} = - \frac{1}{4}\,\uplambda_{\mySM}\,\PH\,G^a_{\mu\nu}\,G^a_{\mu\nu},
\eq
%--
where $\uplambda_{\mySM}$ has inverse mass dimension. The important point is 
that $\uplambda_{\mySM}$ is computed by matching the effective theory to the full 
SM~\cite{Chetyrkin:1997iv,Catani:2001ic,Harlander:2001is,Harlander:2002wh,Harlander:2002vv,Harlander:2003ai,Anastasiou:2002yz,Anastasiou:2002wq,Catani:1998sf,Sterman:2002qn,Kniehl:1995tn,Chetyrkin:1997un}. Even more important is the fact that $\uplambda_{\mySM}$ in the effective theory is the 
renormalized one, with its renormalization constant computed to all 
orders~\cite{Spiridonov:1988md}. Therefore, the logical steps are: first renormalization in 
the full theory, then construction of the effective one.

To be more precise we consider a theory with both light and heavy particles; the
Lagrangian is $\Lag(m)$ where $m$ is the mass of the heavy degree of freedom. Next, we
introduce the corresponding $\Lag_{\eff}$, the effective theory valid up to a scale
$\Lambda = m$. We renormalize the two theories, say in the $\overline{\rm{MS}}\,$-scheme
(taking care that loop-integration and heavy limit are operations that do not commute), and
impose matching conditions among renormalized ``light'' $1$PI Green's functions
%--
\bq
\Gamma^{\ssR}_{\full}(\mu) = \Gamma^{\ssR}_{\eff}(\mu),
\qquad \mu \le m.
\eq
%--
For the case where the full theory is the SM and $m = M_{\PH}$ the whole procedure has been
developed in \Bref{Herrero:1994iu}.
%--
\section{Higgs vertices \label{aHv}}
%--
We are now in the position of writing the complete expression for vertices. There are different 
level of implementation and accuracy. We start with LO-inspired accuracy where the SM vertices
are at LO and the tensor structure of the vertices is the same as the LO SM one but every 
coefficient coming from the effective Lagrangian is kept. Next we go to LO-improved accuracy where
extra tensor structures from the effective Lagrangian is included. Finally there is an
NLO-inspired accuracy where the SM components are at NLO but contributions from $d = 6$ operators 
are included only under the constraint that they do not spoil UV-finiteness.
With the introduction of the following tensors we obtain
%--
\bq
T^{\mu\nu} = p^{\mu}_2\,p^{\nu}_1 - \spro{p_1}{p_2}\,\delta^{\mu\nu},
\quad
P^{\mu\nu} = p^{\mu}_1\,p^{\nu}_1 + 2\,p^{\nu}_1\,p^{\mu}_2 + p^{\mu}_2\,p^{\nu}_2,
\quad
E^{\mu\nu} = \ep^{\alpha\beta\mu\nu}\,p_{1\alpha} p_{2\beta}.
\eq
%--
\vspace{1.5cm}
%--
\fbox{\rule[0.4cm]{0.cm}{1.cm}
\begin{minipage}{0.95\textwidth}
\vspace{1.cm}
\begin{picture}(0,0)(0,0)
 \SetScale{0.4}
 \SetWidth{1.8}
%--
\DashLine(0,0)(50,0){3}
\Photon(50,0)(100,50){2}{5}
\Photon(50,0)(100,-50){2}{5}
\GCirc(50,0){5}{0.6}
%--
\Text(0,10)[]{$P$}
\Text(55,20)[]{$\mu\,p_1$}
\Text(55,-20)[]{$\nu\,p_2$}
%--
\end{picture}
%--
\vspace{-1.5cm}
\bqa
\PHb\,\PAb\,\PAb &{}&
    8\,\frac{\OM}{\HSs}\,
     \lpar \sths\,a^1_{\PV} + \cths\,a^2_{\PV} + g\,\cth\,\sth\,a^3_{\PV}\rpar \,T^{\mu\nu} 
\nl
{}&+& 16\,\frac{\OM}{\HSs}\,
     \lpar \sths\,a^1_{\essV} + \cths\,a^2_{\essV} 
      + g\,\cth\,\sth\,a^3_{\essV}\rpar \,E^{\mu\nu}
\eqa
%--
\vspace{0.1cm}
%--
\end{minipage}}
%--
\vspace{1.cm}
%--

\fbox{\rule[0.4cm]{0.cm}{1.cm}
\begin{minipage}{0.95\textwidth}
\vspace{1.2cm}
\begin{picture}(0,0)(0,0)
 \SetScale{0.4}
 \SetWidth{1.8}
%--
\DashLine(0,0)(50,0){3}
\Line(50,0)(100,50)
\Line(50,0)(100,-50)
\GCirc(50,0){5}{0.6}
%--
\Text(0,10)[]{$P$}
\Text(55,20)[]{$\mu\,p_1$}
\Text(55,-20)[]{$\nu\,p_2$}
%--
\end{picture}
%--
\vspace{-1.5cm}
%--
\bqa
\PHb\,\PZb\,\PZb &{}&
   - g\,\frac{\OM}{\cths}\,\Bigl[ 1 - \lpar 2\,a^1_{\PK} + a^3_{\PK} - 2\,a_{\dPK}\rpar \,
     \ACf \Bigr]\,\delta^{\mu\nu}
\nl
{}&+& \frac{\OM}{\HSs}\,
     \frac{\OMHs}{\cth}\,\lpar \sth\,a^5_{\PK} - \cth\,a^4_{\PK}\rpar \,\delta^{\mu\nu}
\nl
{}&+& 8\,\frac{\OM}{\HSs}\,
     \lpar \cths\,a^1_{\PV} + \sths\,a^2_{\PV}  - g\,\cth\,\sth\,a^3_{\PV}\rpar \,T^{\mu\nu} 
\nl
{}&+& 16\,\frac{\OM}{\HSs}\,
     \lpar \cths\,a^1_{\essV} + \sths\,a^2_{\essV} 
      - g\,\cth\,\sth\,a^3_{\essV} \rpar \,E^{\mu\nu}
\nl
{}&+& \frac{\OM}{\HSs\,\cth}\,\lpar \sth\,a^5_{\PK} - \cth\,a^4_{\PK}\rpar \,P^{\mu\nu}
\eqa
%--
\end{minipage}}
%--
\vspace{1.cm}
%--

\fbox{\rule[0.4cm]{0.cm}{1.cm}
\begin{minipage}{0.95\textwidth}
\vspace{1.2cm}
\begin{picture}(0,0)(0,0)
 \SetScale{0.4}
 \SetWidth{1.8}
%--
\DashLine(0,0)(50,0){3}
\Photon(50,0)(100,50){2}{5}
\Line(50,0)(100,-50)
\GCirc(50,0){5}{0.6}
%--
\Text(0,10)[]{$P$}
\Text(55,20)[]{$\mu\,p_1$}
\Text(55,-20)[]{$\nu\,p_2$}
%--
\end{picture}
%--
\vspace{-1.5cm}
\bqa
\PHb\,\PAb\,\PZb &{}&
    \Bigl[ 4\,g\,\cth\,\lpar 1 - 2\,\sths\rpar \,a^3_{\PV} 
     + 8\,\cths\,\sth\,\lpar a^1_{\PV} - a^2_{\PV}\rpar  
\nl
{}&-& \lpar \cth\,a^5_{\PK} 
     + \sth\,a^4_{\PK}\rpar\,\frac{\OM}{\HSs\,\cth}\Bigr]\,T^{\mu\nu}
\nl
{}&+& 8\,\frac{\OM}{\HSs}\,
     \Bigl[ g\,\lpar 1 - 2\,\sths\rpar \,a^3_{\essV} + 2\,\cth\,\sth\,
     \lpar a^1_{\essV}-a^2_{\essV}\rpar \Bigr]\,E^{\mu\nu}
\nl
{}&-& \frac{\OM}{\HSs\,\cth}\,\lpar \cth\,a^5_{\PK} 
     + \sth\,a^4_{\PK}\rpar \,p^{\mu}_1\,p^{\nu}_1 
\eqa
%--
\end{minipage}}
%--
\vspace{1.cm}

\fbox{\rule[0.4cm]{0.cm}{1.cm}
\begin{minipage}{0.95\textwidth}
\vspace{1.2cm}
\begin{picture}(0,0)(0,0)
 \SetScale{0.4}
 \SetWidth{1.8}
%--
\DashLine(0,0)(50,0){3}
\Line(50,0)(100,50)
\Line(50,0)(100,-50)
\GCirc(50,0){5}{0.6}
%--
\Text(0,10)[]{$P$}
\Text(55,20)[]{$\mu\,p_1$}
\Text(55,-20)[]{$\nu\,p_2$}
%--
\end{picture}
%--
\vspace{-1.5cm}
%--
\bqa
\PHb\,\PWpb\,\PWmb &{}&
    - g\,\OM\,\Bigl[ 1 - \lpar 2\,a^1_{\PK} - a^3_{\PK} - 2\,a_{\dPK}\rpar \,
    \ACf \Bigr]\,\delta^{\mu\nu}
\nl
{}&+& 8\,\frac{\OM}{\HSs}\,a^1_{\PV}\,T^{\mu\nu} 
+ \frac{\OM}{\HSs}\,a^4_{\PK}\,\lpar \spro{P}{P}\,\delta^{\mu\nu} - P^{\mu\nu}\rpar
\nl
{}&+& 16\,\frac{\OM}{\HSs}\,a^1_{\essV}\,E^{\mu\nu} 
\eqa
%--
\end{minipage}}
%--
\vspace{1.cm}

\fbox{\rule[0.4cm]{0.cm}{1.cm}
\begin{minipage}{0.95\textwidth}
\vspace{1.cm}
\begin{picture}(0,0)(0,0)
 \SetScale{0.4}
 \SetWidth{1.8}
%--
\DashLine(0,0)(50,0){3}
\Gluon(50,0)(100,50){2}{5}
\Gluon(50,0)(100,-50){2}{5}
\GCirc(50,0){5}{0.6}
%--
\Text(0,10)[]{$P$}
\Text(55,20)[]{$\mu\,a\,p_1$}
\Text(55,-20)[]{$\nu\,b\,p_2$}
%--
\end{picture}
%--
\vspace{-0.5cm}
\bqa
\PHb\,\Pgb\,\Pgb \quad &{}& \quad
     8\,\frac{\OM}{\HSs}\,a_{\Pg}\,\delta^{a,b}\,T^{\mu\nu}
\eqa
%--
\vspace{0.2cm}
\end{minipage}}
%---
\vspace{1.5cm}

\fbox{\rule[0.4cm]{0.cm}{1.cm}
\begin{minipage}{0.95\textwidth}
\vspace{1.cm}
\begin{picture}(0,0)(0,0)
 \SetScale{0.4}
 \SetWidth{1.8}
%--
\DashLine(0,0)(50,0){3}
\ArrowLine(100,50)(50,0)
\ArrowLine(50,0)(100,-50)
\GCirc(50,0){5}{0.6}
%--
\Text(0,10)[]{$P$}
\Text(55,20)[]{$\mu\,p_1$}
\Text(55,-20)[]{$\nu\,p_2$}
%--
\end{picture}
%--
\vspace{-1.5cm}
%--
\bqa
\PHb\,\PAQtb\,\PQtb &{}&
    - \frac{1}{2}\,g\,\frac{\mt}{\OM}\,
      \Bigl[ 1 + \lpar a^3_{\PK} + 2\,a_{\dPK}\rpar \,\ACf \Bigr]
\nl
{}&+& 2\,\srt\,\ACf \,a^1_{\Pf} 
+ \frac{1}{4\,srt}\,\frac{P^2}{\HSs}\,a^3_{\Pf}
\eqa
%--
\vspace{0.1cm}
%--
\end{minipage}}
%--
\vspace{1.cm}

Similarly for $\Pf = \PQb$ we have
%--
\bqa
\PH(P) \to \PAQb(p_1) + \PQb(p_2) &=&
    - \frac{1}{2}\,g\,\frac{\mb}{\OM}\,
      \Bigl[ 1 + \lpar a^3_{\PK} + 2\,a_{\dPK}\rpar \,\ACf \Bigr]
\nl
{}&+& 2\,\srt\,\ACf \,a^2_{\Pf} 
+ \frac{1}{4\,\srt}\,\frac{P^2}{\HSs}\,a^4_{\Pf}
\eqa
%--
\section{$\PZ$ couplings \label{aZv}}
%--
The $\PZ \PAf \Pf$ vertex can be parametrized as follows:
%--
\bq
\frac{i g}{2 \cth}\,\uprho_{\Pf}\,\gamma^{\mu}\,\Bigl[ I_{3\Pf}\,(1 + \gamma^5) -
2\,Q_{\Pf}\,\kappa_{\Pf}\,\sths \Bigr],
\eq
%--
where $I_{3\Pf}$ is the third component of isospin and $Q_l = -1$, $Q_{\nu}= 0$,
$Q_u= 2/3$ and $Q_d= -1/3$. The anomalous part reads as follows;
%--
\bqa
\Delta\uprho_{\Pf} &=& \ACf \,\Bigl[ a^3_{\PK} - 32\,Q_{\Pf}\,I_{3\Pf}
     \lpar 1 - \sth\rpar)\,\cth^3\,a^3_{\PV} \Bigr],
\nl
\Delta\kappa_{\Pf} &=& 2\,\ACf \,\frac{\cths}{\sths}\,\Bigl[
     a^3_{\PK} + 4\,\lpar 1 + 4\,Q_{\Pf}\,I_{3\Pf}\,\sth(1-\sth)\rpar\,\sth\cth\,a^3_{\PV} \Bigr]
\eqa
%--
\section{Partial decay widths \label{pdw}}
\begin{figure}[h]
\vspace{1.cm}
\begin{picture}(0,0)(-150,0)
 \SetScale{0.4}
 \SetWidth{1.8}
%--
\DashLine(0,0)(100,0){3}
\Line(100,0)(150,50)
\Line(100,0)(150,-50)
\GCirc(100,0){15}{1}
\GCirc(0,0){5}{0}
%--
\DashLine(200,0)(300,0){3}
\Line(300,0)(350,50)
\Line(300,0)(350,-50)
\GCirc(300,0){7}{0.6}
%--
\Text(-20,0)[]{\Large $\sum_i$}
\Text(40,0)[]{i}
\Text(65,0)[]{\Large $+$}
%--
\end{picture}
\vspace{1.5cm}
\caption[]{Amplitude for a two-body decay of the Higgs boson (dash line) including LO+NLO SM 
contributions with a sum over all one-loop diagrams (i); SM diagrams are eventually multiplied 
by a universal scaling from $d = 6$ operators (black circle); the grey circle represents
a contact term.}
\label{setupPW}
\end{figure}
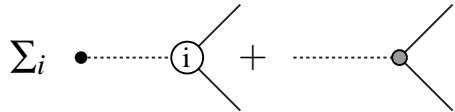

%--
In this Section we compute the partial decay widths of the Higgs boson for the most relevant
channels: first we introduces the dimensionless coupling
%--
\bq
g_6 = \frac{1}{\myGF\,\HSs} = 0.085736\,\lpar \frac{\UTeV}{\Lambda}\rpar^2
\eq
%--
which parametrizes deviations from the SM results. Furthermore, we introduce new couplings
%--
\bq
g\,a^1_{\PV} = A^1_{\PV}, \quad g\,a^2_{\PV} = A^2_{\PV}, \quad 
g^2\,a^3_{\PV} = A^3_{\PV}, \quad g\,a_{\Pg} = A_{\Pg} 
\eq
%--
\bq
g^2\,a^1_{\PK} = A^1_{\PK}, \quad g^2\,a^3_{\PK} = A^3_{\PK}, \quad
g^2\,a_{\dPK} = A_{\dPK}, 
\eq
%--
\bq
g\,a^1_{\Pf} = \frac{1}{4\,\srt}\,\frac{\mt}{\OM}\,A^1_{\Pf}, \qquad 
g\,a^2_{\Pf} = \frac{1}{4\,\srt}\,\frac{\mb}{\OM}\,A^2_{\Pf},
\eq
%--
and express all amplitudes in terms of a SM-component (eventually scaled by the effect of
$d = 6$ operators) and by a contact component, as shown in \refF{setupPW}.
We introduce an auxiliary coefficient,
%--
\bq
A^0_{\PK} = A^1_{\PK} + 2\,\frac{A^3_{\PK}}{\sths} + 4\,A_{\dPK}.
\eq
%--
%--

We will now show results for various decay processes.

\begin{itemize}
\item{{\underline{$\PHb \to \PGgb \PGgb$}}}
\end{itemize}
%--
For $\PH \to \PGg \PGg$ the SM amplitude reads
%--
\bq
\Amp_{\mySM} = F_{\mySM}\,\lpar \delta^{\mu\nu} + 2\,\frac{p^{\nu}_1 p^{\mu}_2}{\OMHs}\rpar\,
e_{\mu}\lpar p_1\rpar\,e_{\nu}\lpar p_2\rpar\,
\eq
%--
where
%--
\bq
F_{\mySM} = -g\,\OM\,F^{\PW}_{\mySM} 
- \frac{1}{2}\,g\,\frac{\mt^2}{\OM}\,F^{\PQt}_{\mySM}
- \frac{1}{2}\,g\,\frac{\mb^2}{\OM}\,F^{\PQb}_{\mySM}.
\eq
%--
\bqa
F^{\PW}_{\mySM} &=& 6 + \frac{\OMHs}{\OMs} + 6\,\lpar \OMHs - 2\,\OMs\rpar\, 
C_0\lpar -\OMHs\,,\,0\,,\,0\,;\,\OM\,,\OM\,,\OM\rpar,
\nl
F^{\PQt}_{\mySM} &=& - 8 - 4\,\lpar \OMHs - 4\,\mt^2\rpar\,
C_0\lpar -\OMHs\,,\,0\,,\,0\,;\,\mt\,,\mt\,,\mt\rpar,
\eqa
%--
\etc The $C_0$ function is given by
%--
\bq
C_0\lpar - M^2\,,\,0\,,\,0\,;\,m\,,m\,,m\rpar = - \frac{1}{2\,M^2}\,
\ln^2\,\frac{(1 - x)^{1/2} + 1}{(1 - x)^{1/2} - 1},
\eq
%--
where $x = 4\,(m^2 - i\,0)/M^2$. Note that there is no need to split the result for this
$C_0\,$-function into the two regions $x > 1$ and $x \le 1$ since the $i\,0$ prescription
uniquely defines the analytic continuation. We find
%--
\bqa
\Amp_{\PH \to \PGg\PGg} &=&
   \lpar 4\,\srt\,\myGF\rpar^{1/2}\,\Bigl\{ -\,\frac{\alpha}{\pi}\,\Bigl[
C^{\PGg\PGg}_{\PW}\,F^{\PW}_{\mySM} + 3\,Q^2_{\PQt}\,C^{\PGg\PGg}_{\PQt}\,F^{\PQt}_{\mySM} + 
3\,Q^2_{\PQb}\,C^{\PGg\PGg}_{\PQb}\,F^{\PQb}_{\mySM} \Bigr] + F_{\AC} \Bigr\}
\nl
F_{\AC} &=& \frac{g_6}{\srt}\,\OMHs\,
       \lpar  \sths\,A^1_{\PV} + \cths\,A^2_{\PV} + \cth\sth\,A^3_{\PV} \rpar.
\label{ggAC}
\eqa
%--
where the scaling factors are given by
%--
\bq
C^{\PGg\PGg}_{\PW} = 
 \frac{1}{4}\,\OMs \Bigl\{ 
  1 + \frac{g_6}{4\,\srt}\,
   \Bigl[ 8\,A^3_{\PV}\,\cth\,\lpar \sth + \frac{1}{\sth}\rpar + A^0_{\PK} \Bigr] \Bigr\}
\label{CV}
\eq
%--
for the $\PW\,$-loop and
%--
\bq
C^{\PGg\PGg}_{\PQt} = 
  \frac{1}{8}\,\mts\,\Bigl\{
  1 + \frac{g_6}{4\,\srt}\,
   \Bigr[ 8\,A^3_{\PV}\,\cth\,\lpar \sth + \frac{1}{\sth}\rpar + A^0_{\PK} - A^1_{\Pf} 
   \Bigr] \Bigr\}
\label{Cf}
\eq
%--
\bq
C^{\PGg\PGg}_{\PQb} = 
  \frac{1}{8}\,\mbs\,\Bigl\{
  1 + \frac{g_6}{4\,\srt}\,
   \Bigl[ 8\,A^3_{\PV}\,\cth\,\lpar \sth + \frac{1}{\sth}\rpar + A^0_{\PK} - A^2_{\Pf} 
   \Bigr]\Bigr\}
\eq
%--
for the quark loops.

The amplitude is the sum of the $\PW$, $\PQt$ and $\PQb$ SM components, each scaled
by some combination of Wilson coefficients, and of a {\em contact} term. The latter
is $\ord{g_6}$ while the rest of the corrections is $\ord{\frac{\alpha}{\pi}\,g_6)}$.
However, one should remember that $O^i_{\PV}$ are operators of $L\,$-type, \ie they arise 
from loop correction in the complete theory. Therefore, the corresponding coefficients
are expected to be very small although this is only an argument about naturalness without
a specific quantitative counterpart (a part from a $1/(16\,\pi^2)$ factor from loop integration).

The result for $\PH \to \Pg \Pg$ follows straightforwardly.
%--
\begin{itemize}
\item{\underline{$\PHb \to \Pgb \Pgb$}}
\end{itemize}

The result for $\PH \to \Pg \Pg$ is straightforward. Including also the $\PQb\,$-loop we obtain
%--
\bq
\Amp_{\PH \to \Pg\Pg} =
   \lpar 4\,\srt\,\myGF\rpar^{1/2}\,\Bigl[ -\,\frac{\alphas(\OMHs)}{\pi}\,\lpar
C^{\Pg\Pg}_{\PQt}\,F^{\PQt}_{\mySM} + 
C^{\Pg\Pg}_{\PQb}\,F^{\PQb}_{\mySM} \rpar
 + \frac{g_6}{\srt}\,\OMHs\,A_{\Pg} \Bigr]
\eq
%--
where the scaling of the quark components is given by
%--
\bq
C^{\Pg\Pg}_{\PQt} = \frac{1}{16}\,\mt^2\,\Bigl[ 1 
+ \frac{g_6}{4\,\srt}\,\lpar
 A^0_{\PK} - A^1_{\Pf}\rpar \Bigr]
\eq
%--
\bq
C^{\Pg\Pg}_{\PQb} = \frac{1}{16}\,\mb^2\,\Bigl[ 1 
+ \frac{g_6}{4\,\srt}\,\lpar
 A^0_{\PK} - A^2_{\Pf}\rpar \Bigr]
\eq
%--
\begin{itemize}
\item{\underline{$\PHb \to \PAQbb \PQbb$}}
\end{itemize}
%--
For the $\PH \to \PAQb \PQb$ amplitude we have to examine again if the are UV-scalable
diagrams. In this case there is a tree-level amplitude, and renormalization is required.
At NLO there are two type of diagrams, the abelian ones involving the $\PH \mybar{\Pf} \Pf$
vertex and the non-abelian ones involving a $\PH \PV \PV$ ($\PH \PV \upphi, \PH \upphi \upphi$) 
vertex. Therefore we have to search for the unique combination that multiply all the vertices, 
which is
%--
\bq
4\,\ACf \,a_{\dPK}.
\eq
%--
The SM amplitude reads as follows:
%--
\bq
\Amp^{\mySM}_{\PH \to \PAQb \PQb} = g^3\,\frac{\mb}{\OM}\,F^{\mySM}_{\PH \to \PAQb \PQb}\,
\mybar{u}(p_2)\,v(p_1).
\eq
%--
The expression for $F^{\mySM}_{\PH \to \PAQb \PQb}$ can be found in Section 5.9.4 of 
\Bref{Bardin:1999ak}. Renormalization and QCD corrections are discussed in Section 
(11.2${-}$11.4) of \Bref{Bardin:1999ak}.
The complete amplitude for $\PH \to \PAQb \PQb$ is
%--
\bqa
\Amp_{\PH \to \PAQb \PQb} &=& \lpar 4\,\srt\,\myGF\rpar^{1/2}\,\mb\,\mybar{u}(p_2)\,v(p_1)
\Bigl\{
\frac{\myGF\,\OMs}{\pi^2}\,\,C^{\PAQb\PQb}\,
F^{\mySM}_{\PH \to \PAQb \PQb}
\nl
{}&+& \frac{g_6}{128\,\srt}\,\Bigl[
\frac{\OMHs}{\OMs}\,A^4_{\Pf} - 16\,
\lpar A^3_{\PK} + 2\,A_{\dPK} + A^2_{\Pf} \rpar\Bigr]\Bigr\},
\eqa
%--
\bq
C^{\PAQb\PQb} = \frac{1}{2\,\srt}\Bigl[1  + \frac{g_6}{4\,\srt}\,
\lpar A^1_{\PK} + A^3_{\PK} + 6\,A_{\dPK}\rpar \Bigr].
\eq
%--
In the SM, NLO corrections to the amplitude include a QED part so that, technically speaking,
the process is $\PH \to \PAQb \PQb(\PGg)$, \ie real corrections are added. There is also
a contribution from the $d = 6$ operators
%--
\bq
\frac{i g}{3\,\srt\,\HSs}\,\sth\,A^4_{\Pf}\,\mybar{b}\,\gamma^5\,b\,\myPAB_{\mu}\,
\pdmu\HB,
\eq
%--
which, however, is not infrared divergent and will not be included.

In this Section we have considered partial decay widths of the SM Higgs boson; in the SM, the 
common belief is that (for a light Higgs boson) the product of on-shell production 
cross-section (say in gluon-gluon fusion) and branching ratios (zero-width approximation or 
ZWA) reproduces the correct result to great accuracy. The work of \Bref{Kauer:2012hd} shows the 
inadequacy of ZWA for a light Higgs boson signal at the level of $5\%$. Therefore, one
should always implement the results of this Section within a consistent off-shell formulation
of the problem. 
%--
\subsection{$\Pgb\Pgb \to \PHb$}
%--
In ZWA the inclusive cross section for the production of the SM Higgs boson in hadronic 
collisions can be written as
%--
\bqa 
  \sigma \lpar s,\mhs \rpar &=&
  \sum_{i,j} \, \int_0^1 \! dx_1  \int_0^1 \! dx_2 \,\,
  f_{i / h_1}\lpar x_1,\Fsc \rpar \, 
  f_{j / h_2}\lpar x_2,\Fsc \rpar \,
  \times \nl {}&\times&
  \int_0^1 \! dz \, \delta \lpar z -\frac{\mhs}{s\, x_1 x_2} \rpar 
  \,z\, \sigma^{(0)}\, 
  G_{ij}\lpar z;\alphas(\Rsc),\mhs/\Rsc; \mhs/ \Fsc \rpar,
\label{eq:CShad}
\eqa
%--
where $\sqrt{s}$ is the center-of-mass energy and $\mu_{\ssF}$ and 
$\mu_{\ssR}$ stand for factorization and renormalization scales.

In \eqn{eq:CShad} the partonic cross section for the sub-process $ij\to H+X$, 
with $i(j) = \Pg, \PQq, \PAQq$, has been convoluted with the parton 
densities $f_{a/h_b}$ for the colliding hadrons $h_1$ and $h_2$. The Born 
factor $\sigma^{(0)}$ reads
%--
\bq
\sigma^{(0)} =  \frac{\myGF}{288\srt\pi} 
                \left| \sum_{\PQq= \PQt,\PQb}\,\Amp^{\mySM}_{\PQq} \right|^2,
\label{eq:oth}
\eq
%--
where $\myGF$ is the Fermi-coupling constant; the amplitude is generalized to
%--
\bq
\Amp = \sum_{\PQq= \PQt,\PQb}\,c^{\Pg\Pg}_{\PQq}\,\Amp^{\mySM}_{\PQq} + \Amp_{\AC},
\label{imp_QCD}
\eq
%--
where the last term is induced by the operator $\Ope_{\Pg}$, and where the scaling factors are
%--
\bq
c^{\Pg\Pg}_{\PQt} = 1 + \frac{g_6}{4\,\srt}\,\lpar
% A^1_{\PK} + 2\,A^3_{\PK} + 4\,A_{\dPK} - A^1_{\Pf}\rpar 
 A^0_{\PK} - A^1_{\Pf}\rpar 
%\eq
%--
%\bq
\qquad
%--
c^{\Pg\Pg}_{\PQb} = 1 + \frac{g_6}{4\,\srt}\,\lpar
% A^1_{\PK} + 2\,A^3_{\PK} + 4\,A_{\dPK} - A^2_{\Pf}\rpar.
 A^0_{\PK} - A^2_{\Pf}\rpar.
\eq
%--
Since $\Amp^{\mySM}_{\PQt}$ and $\Amp^{\mySM}_{\PQb}$ are separately UV finite it is possible
to include NLO(NNLO) QCD corrections even in presence of anomalous scaling factors
The coefficient functions $G_{ij}$ of \eqn{eq:CShad}can be computed in QCD through a 
perturbative expansion in the strong-coupling constant $\alpha_\ssS$,
%--
\bq
G_{ij} \lpar z ; \alphas(\Rsc) , \mhs/\Rsc ; \mhs/\Fsc \rpar = \alphas^2(\Rsc) 
       \sum_{n=0}^{\infty} \lpar \frac{\alphas(\Rsc)}{\pi}\rpar^n
       G_{ij}^{(n)}\lpar z;\mhs/\Rsc;\mhs/\Fsc \rpar,
\eq
%--
with a scale-independent LO contribution given by
%--
\bq
G^{(0)}_{ij}(z) = \delta_{ig}\,\delta_{jg}\,\delta\lpar 1 - z\rpar.
\eq
%--
The NLO QCD coefficients have been computed in Ref.~\cite{Spira:1995rr}, 
keeping the exact $\mt$ and $\mb$ dependence. NNLO results have been
derived in Ref.~\cite{Harlander:2000mg} in the large $\mt$ limit (see
\Bref{Dawson:1991zj} for the NLO case); analytical expressions can be found in 
\Bref{Anastasiou:2002yz}. 
The accuracy of these fixed-order computations has been improved with soft-gluon
resummed calculations~\cite{Catani:2003zt,Moch:2005ky,Laenen:2005uz}.

QCD corrections cannot be implemented in the additive part of \eqn{imp_QCD}. To do that one needs 
a model for $\Ope_{\Pg}$, as done in Section~2 of \Bref{Bonciani:2007ex} where the SM is extended
to included colored scalars, so that one has
%--
\bq
\Amp = \sum_{\PQq}\,\Amp_{\PQq} + \sum_{\PS}\,\Amp_{\PS},
\eq
%--
where fermions and scalars transform according to some $\lpar C\,,\,T\,,\,Y\rpar$ representation of
$SU(3)\,\otimes\,SU(2)\,\otimes\,U(1)$, as long as ${\mybar{C}}\,\otimes\,C \;\ni\; 8$.
Complete QCD corrections for fermion and scalar amplitudes have been computed in 
\Bref{Bonciani:2007ex}.
%--
\subsection{Simplified scenario}
%--
If we restrict the scenario to bosonic $T\,$-operators only ($a_{\PK}$, $a^{1,3}_{\PK}$ and
$a_{\dPK}$) the scaling factors are:
%--
\bq
C^{\PGg\PGg}_{\PW} = C^{\PGg\PGg}_{\PQt} = C^{\PGg\PGg}_{\PQb} =
 \frac{1}{4}\,\OMs \Bigl[ 
  1 + \frac{g_6}{4\,\srt}\,
   A^0_{\PK} \Bigr]
\eq
%--
\bq
C^{\Pg\Pg}_{\PQt} = \frac{1}{16}\,\mt^2\,\Bigl[ 1 
+ \frac{g_6}{4\,\srt}\,
A^0_{\PK} \Bigr]
\qquad
C^{\Pg\Pg}_{\PQb} = \frac{1}{16}\,\mb^2\,\Bigl[ 1 
+ \frac{g_6}{4\,\srt}\,
A^0_{\PK} \Bigr]
\eq
%--
\bq
C^{\PAQb\PQb} = \frac{1}{2\,\srt}\,\Bigl[ 1 + \frac{g_6}{4\,\srt}\,
\lpar A^1_{\PK} + A^3_{\PK} + 6\,A_{\dPK}\rpar\Bigr].
\eq
%--
The {\em contact} terms are all zero but
%--
\bq
\Amp^{\rm ct}_{\PH \to \PAQb \PQb} =
 -\,\frac{g_6}{16\,\srt}\,\mb\,\lpar A^3_{\PK} + 2\,A_{\dPK} \rpar.
\eq
%--
In this case it is not possible to differentiate bosonic loops from quark loops.
%--
\subsection{BSM Lagrangians}
%--
By BSM Lagrangians we mean those Lagrangians containing new, heavy degrees of freedom that
can produce $d= 6$ operators when the heavy particles are integrated out.
One of the most important questions is about the sign of the Wilson coefficients $a_i$ in 
\eqn{FLag}, \ie to find the set of coefficients such that
%--
\bq
\{ a_i \mid a_i > 0 \}  \quad \in \quad \{ \Lag_{+} \}.
\eq
%--
Before entering the discussion on BSM Lagrangian we recall few, well-know, facts about tree-level
custodial symmetry. The SM Higgs potential is invariant under $SO(4)$; furthermore, 
$SO(4) \sim SU(2)_{\ssL}\,\otimes\,SU(2)_{\ssR}$ and the Higgs VEV breaks it down to the diagonal 
subgroup $SU(2)_{\ssV}$. It is an approximate symmetry since the $U(1)_{\ssY}$ is a subgroup of 
$SU(2)_{\ssR}$ and only that subgroup is gauged. Furthermore, the Yukawa interactions are only 
invariant under $SU(2)_{\ssL}\,\otimes\,U(1)_{\ssY}$ and not under 
$SU(2)_{\ssL}\,\otimes\,SU(2)_{\ssR}$ and therefore not under the custodial subgroup.
Therefore, if we require a new CP-even scalar, which is also in a custodial representation of
the group, the $\PW/\PZ\,$-bosons can only couple to a singlet or a $5\,$-plet, as discussed
in \Bref{Low:2012rj}. If $(N_{\ssL}\,,\,N_{\ssR})$ denotes a representation of 
$SU(2)_{\ssL}\,\otimes\,SU(2)_{\ssR}$, the usual Higgs doublet scalar is a $(2\,,\,\mybar{2})$, 
while the $(3\,,\,\mybar{3}) = 1\,\oplus\,3\,\oplus\,5$ contains the Higgs-Kibble ghosts (the $3$),
a real triplet (with $Y = 2$) and a complex triplet (with $Y = 0$). The 
Georgi - Machaceck model~\cite{Georgi:1985nv} has EWSB from both a $(2\,,\,\mybar{2})$ and 
a $(3\,,\,\mybar{3})$.

To introduce the discussion on BSM Lagrangians we define the following quantity:
%--
\bq
\Delta C= g_6\,A^0_{\PK}.
\eq
%--
Assuming $A^3_{\PV} = 0$ and requiring that the coupling $\PH\PWp\PWm$ in the decay $\PH \to
\PGg\PGg$ has the standard value, \ie that
%--
\bq
C^{\PGg\PGg}_{\PW} = C^{\PGg\PGg}_{\PW}\,\bmid_{\mySM}
\label{SMS}
\eq
%--
we obtain the condition $\Delta C = 0$. We now examine different models and explicitly compute 
the corresponding value for $\Delta C$. At the same time we address the question of models
allowing for non-standard coupling $\PH\PAQt\PQt$ in the loop for $\PH \to \PGg\PGg$. 

In general, the basis for a representation of $SU(2)$ can be characterized~\cite{Einhorn:1981cy}
as a tensor field
%--
\bq
\uppsi_{i_1\,\cdots\,i_n} \to G_{i_1\,j_1}\,\cdots\,G_{i_n\,j_n}\,\uppsi_{j_1\,\cdots\,j_n},
\eq
%--
where $G$ are $SU(2)\,$-matrices. An irreducible representation of spin $n/2$ is characterized
by a totally symmetric field with $n$ indices. The hermitian conjugate 
$\uppsi^{\dagger}_{i_1\,\cdots\,i_n}$ transforms according to the complex conjugate representation,
$\uppsi^{i_1\,\cdots\,i_n}$ and indices can be lowered using the metric tensor
$e_{ij}$. To define the covariant derivative we introduce
%--
\bq
\PI^{i'_1\,\cdots\,i'_n}_{i_1\,\cdots\,i_n} = \prod_{r=1}^n\,\delta^{i'_r}_{i_r} 
\eq
%--
\bq
D^{i'_1\,\dots\,{\hat i}'_l\,\dots\,i'_n}_{i_1\,\dots\,{\hat i}_l\,\dots\,i_n} = 
\prod_{r=1}^{l-1}\,\delta^{i'_r}_{i_r}\,\lpar -\,\frac{i}{2}\,\uptau_a \rpar^{i'_l}_{i_l}\,
\prod_{s=l+1}^{n}\,\delta^{i'_s}_{i_s},
\qquad
U^{i'_1\,\dots\,{\hat i}'_l\,\dots\,i'_n}_{i_1\,\dots\,{\hat i}_l\,\dots\,i_n} = 
\prod_{r=1}^{l-1}\,\delta^{i'_r}_{i_r}\,\lpar \frac{i}{2}\,\uptau_a \rpar^{i'_l}_{i_l}\,
\prod_{s=l+1}^{n}\,\delta^{i'_s}_{i_s}.
\eq
%--
The covariant derivative is
%--
\bqa
\lpar D_{\mu}\,\uppsi\rpar^{j_1\,\dots\,j_n}_{i_1\,\dots\,i_n} &=&
\Bigl\{
\PI^{i'_1\,\cdots\,i'_n}_{i_1\,\cdots\,i_n}\,\PI^{j_1\,\cdots\,j_n}_{j'_1\,\cdots\,j'_n}\,
\partial_{\mu}
+ g\,\PW^a_{\mu}\,\PI^{j_1\,\cdots\,j_n}_{j'_1\,\cdots\,j'_n}\,\sum_{l=1}^n\,
D^{i'_1\,\dots\,{\hat i}'_l\,\dots\,i'_n}_{i_1\,\dots\,{\hat i}_l\,\dots\,i_n} 
\nl
{}&+& g\,\PW^a_{\mu}\,\PI^{i'_1\,\cdots\,i'_n}_{i_1\,\cdots\,i_n}\,\sum_{l=1}^n\,
U^{j_1\,\dots\,{\hat j}_l\,\dots\,j_n}_{j'_1\,\dots\,{\hat j}'_l\,\dots\,j_n}\Bigr\}\, 
\uppsi^{j'_1\,\dots\,j'_n}_{i'_1\,\dots\,i'_n},
\eqa
%--
where $a=0,\dots,3$, $\PW^{1,2,3}_{\mu} = B^{1,2,3}_{\mu}$ and $\PW^0_{\mu}= g_1\,B^0_{\mu}$. 

Here are a few examples of BSM Lagrangians.

%--
\begin{itemize}
\item{\bf{Example} $\mathbf{1}$}
\end{itemize}
%--
Consider the following Lagrangian~\cite{Bonnet:2011yx}:
%--
\bq
\Lag_1 = \Lag_{\mySM} + \Lag_s
\eq
%--
\bqa
\Lag_s &=&
       -\,\frac{1}{2} \, \pdmu \PS \pdmu \PS
       - \frac{1}{2}\,\mPs\, \PS^2
       + \mu_{\PS}\,\PKdag \PK \PS
       -  \,\frac{1}{2} \, \lpar D_{\mu} \upeta\rpar^a \lpar D_{\mu} \upeta\rpar^a
       - \frac{1}{2}\,M^2_{\upeta}\, \upeta^a\upeta^a
\nl
{}&+& \mu_{\ssT}\,\PKdag \uptau_a \PK \upeta^a
       -  \,\lpar D_{\mu} \upxi\rpar^{\dagger\,a} \lpar D_{\mu} \upxi\rpar^a
       - M^2_{\upxi}\, \upxi^{\dagger a} \upeta^a
       + \Bigl[ \mu_{\upxi}\,\PKdag \uptau_a \PK^c \upxi^a + \mbox{h.c.}\Bigr]
\eqa
%--
where $\PS$ is a scalar singlet and $\upeta,\upxi$ are scalar triplets with different 
hypercharge, see \Brefs{Ross:1975fq,Passarino:1989py}.
To be more precise, $\upeta$ can be written as a complex symmetric tensor of rank two
and $\upxi$ as a traceless tensor. In our case we introduce
%--
\bq
\PX_{\PS} = \frac{\mu^2_{\PS}}{\myGF\,M^4_{\PS}},
\quad
\PX_{\upeta} = \frac{\mu^2_{\upeta}}{\myGF\,M^4_{\upeta}},
\quad
\PX_{\upxi} = \frac{\mid \mu_{\upxi}\mid^2}{\myGF\,M^4_{\upeta}}.
\eq
%--
Projecting onto the $d= 6$ operators we obtain
%--
\bq
\Delta C = - 2\,\Bigl[ 
\lpar \frac{2}{\sths} - 1\rpar\,\PX_{\upeta} 
- 2\,\PX_{\PS} 
- 2\,\,\lpar 1 + \frac{2}{\sths}\rpar\,\PX_{\upxi} \Bigr].
\eq
%--
The scenario of \eqn{SMS} has a solution 
%--
\bq
 \PX_{\PS} = 
 \lpar \frac{1}{\sths} - \frac{1}{2}\rpar\,\PX_{\upeta} 
 - \lpar 1 + \frac{2}{\sths}\rpar\,\PX_{\upxi},
\eq
%--
which requires the condition
%--
\bq
\PX_{\upeta} \ge 2\,\frac{2 + \sths}{2 - \sths}\,\PX_{\upxi}.
\eq
%--
\begin{itemize}
\item{\bf{Example} $\mathbf{2}$}
\end{itemize}
%--
Alternatively, we could consider a Lagrangian~\cite{Bonnet:2011yx} 
%--
\bq
\Lag_2 = \Lag_{\mySM} + \Lag_v
\eq
%--
\bqa
\Lag_v &=&
- \frac{1}{4}\,\PV_{\mu\nu}\,\PV_{\mu\nu} - \frac{1}{2}\,M^2_{\PV}\,\PV_{\mu} \PV_{\mu} -
i\,g_{\PV}\,V_{\mu}\,\Bigl[ \lpar( D_{\mu}\,\PK\rpar^{\dagger}\,\PK -
\PKdag\,D_{\mu}\,\PK\Bigr]
\nl
{}&- &
\frac{1}{4}\,\PU^a_{\mu\nu}\,\PU^aV_{\mu\nu} - \frac{1}{2}\,M^2_{\PU}\,\PV^a_{\mu} \PV^a_{\mu} -
\frac{i}{2}\,g_{\PU}\,V^a_{\mu}\,\Bigl[ \lpar( D_{\mu}\,\PK\rpar^{\dagger}\,\uptau_a\,\PK -
\PKdag\,\uptau_a\,D_{\mu}\,\PK\Bigr],
\eqa
%--
which contains $I= 0$ and $I= 1$ new vector fields; introducing
%--
\bq
\PX_{\PV\,\PU} = \frac{g^2_{\PV,\PU}}{\myGF\,M^2_{\PU,\PV}} 
\eq
%--
we obtain that the scenario of \eqn{SMS} requires 
%--
\bq
\PX_{\PU} = 8\,\frac{\cths}{\sths}\,PX_{\PV}.
\eq
%--
\begin{itemize}
\item{\bf{Example} $\mathbf{3}$}
\end{itemize}
%--
A mixture of vector and scalar fields~\cite{Bonnet:2011yx}, \eg
%--
\bq
\Lag_3 = \Lag_{\mySM} + \Lag_{sv}
\eq
%--
\bqa
\Lag_{sv} &=&
       -\,\frac{1}{2} \, \pdmu \PS \pdmu \PS
       - \frac{1}{2}\,\mPs\, \PS^2
       + \mu_{\PS}\,\PKdag \PK \PS
       - \mu_{\PV\PS}\,\PV_{\mu}\,\pdmu \PS,
\nl
{}&-& \frac{1}{4}\,\PV_{\mu\nu}\,\PV_{\mu\nu} - \frac{1}{2}\,M^2_{\PV}\,\PV_{\mu} \PV_{\mu} -
i\,g_{\PV}\,V_{\mu}\,\Bigl[ \lpar D_{\mu}\,\PK\rpar^{\dagger}\,\PK -
\PKdag\,D_{\mu}\,\PK\Bigr]
\nl
{}&-&
 \frac{1}{4}\,\PU^a_{\mu\nu}\,\PU^aV_{\mu\nu} - \frac{1}{2}\,M^2_{\PU}\,\PV^a_{\mu} \PV^a_{\mu} -
 \frac{i}{2}\,g_{\PU}\,V^a_{\mu}\,\Bigl[ \lpar D_{\mu}\,\PK\rpar^{\dagger}\,\uptau_a\,\PK -
 \PKdag\,\uptau_a\,D_{\mu}\,\PK\Bigr],
\eqa
%--
gives
%--
\bq
\Delta C = 
\frac{1}{2}\,\PX_{\PU}
- 4\,\frac{\cths}{\sths}\,\PX_{\PV}
+ 4\,\PX_{\PS}\,\lpar 1 - \frac{\mu^2_{\PV\PS}}{M^2_{\PV}}\rpar. 
\eq
%--
The scenario of \eqn{SMS} requires large values for $\mu_{\PV\PS}$. When we include all scalar
an vector fields, \eqn{SMS} is satisfied by
%--
\bq
\PX_{\PU} = 4\,\Bigl[ 
2\,\frac{\cths}{\sths}\,\PX_{\PV} +
\PX_{\upeta} - 6\,\PX_{\upxi} - 2\,\PX_{\PS}\,\lpar 1 - 
\frac{\mu^2_{\PV\PS}}{M^2_{\PV}}\rpar\Bigr]
\eq
%--  
\begin{itemize}
\item{\bf{Example} $\mathbf{4}$}
\end{itemize}
%--
In order to differentiate the bosonic amplitude from the fermionic one we need 
$\Ope^{1,2}_{\Pf}$. One way to introduce them is to consider an additional Lagrangian,
%--
\bq
\Lag_4 = \Lag_{\mySM} + \Lag_{\upchi}
\eq
%--
where $\upchi$ is a doublet
%--
\bqa
\Lag_{\upchi} &=&
       -\,\frac{1}{2} \, \lpar D_{\mu} \upchi\rpar^{\dagger} D_{\mu} \upchi
       - \frac{1}{2}\,M^2_{\upchi}\, \upchi^{\dagger} \upchi +
       \Bigl[ \uplambda_{\upchi} \lpar \PKdag \PK\rpar\,\lpar \PKdag \upchi\rpar +
              \mbox{h.c.} \Bigr]
\nl
{}&+& \Bigl[ Y_{\upchi}\,\POpsiL\,\upchi^c\,\PQtR +
      y_{\upchi}\,\POpsiL\,\upchi\,\PQbR + \mbox{h.c.} \Bigr],
\eqa
%--
which would produce $a^1_{\Pf}$ of the order of $(Y_{\upchi}\,\uplambda_{\upchi})/M^2_{\upchi}$.

Finally, we examine the possibility of a non-zero $F_{\AC}$ in \eqn{ggAC}. This requires
$\Ope_{\PV}$ operators. One option is to include colored scalar fields~\cite{Aglietti:2006tp}
but we could also include a real triplet~\cite{Georgi:1985nv,Passarino:1989py,Logan:2010en}
%--
\bq
\Pxid = \lpar \Pxim\,,\,\Pxiz\,,\,\Pxip\rpar,
\eq
%--
with hypercharge $Y = 0$. The Lagrangian reads as follows
%-- 
\bq
\Lag_{\upxi} =
    -  \,\lpar D_{\mu} \Pxi \rpar^{\dagger}\,D_{\mu} \Pxi
    - M^2_{\xi}\, \Pxid \Pxi
    + \uplambda_{\upxi}\,\lpar \PKdag K\rpar\,\lpar \Pxid \Pxi \rpar,
\eq
%--
with covariant derivative $D_{\mu} = \pdmu - i\,g\,B^a_{\mu}\,T_a$ and
%--
\[
T_1 = \frac{1}{\srt}\,\left(
\begin{array}{ccc}
0 & 1 & 0 \\
1 & 0 & 1 \\
0 & 1 & 0 
\end{array}
\right)
%--
\qquad
%--
T_2 = \frac{1}{\srt}\,\left(
\begin{array}{ccc}
0 & -i & 0 \\
i & 0 & -i \\
0 & i & 0 
\end{array}
\right)
%--
\qquad
%--
T_3 = \left(
\begin{array}{ccc}
1 & 0 & 0 \\
0 & 0 & 0 \\
0 & 0 & -1 
\end{array}
\right)
\]
%--
which gives the following couplings:
%--
\bq
i\,\,\stw\,\PA_{\mu}\,\lpar \Pxip \pdmu \Pxim - \Pxim \pdmu \Pxip \rpar
\quad
4\,\uplambda_{\upxi}\,\frac{M}{g}\,\PH \Pxip \Pxim,
\quad
- g^2\,\stws\,\PA_{\mu} \PA_{\mu}\,\Pxip \Pxim
\eq
%--
and produces a loop of $\upxi\,$scalars in the $\PH\PGg\PGg$ coupling.

Additional examples of BSM Lagrangians can be found in \Brefs{Craig:2012vn,Alves:2012yp}
and in \Bref{Yagyu:2012qp}. General studies can also be found in 
\Brefs{ArkaniHamed:2012kq,Carmi:2012in,Espinosa:2012vu}.
%--
\subsection{MSSM}
%--
In this paper we assume that the starting point in comparing theory with data is the SM.
Another choice could be to start from the Minimal-Supersymmetric Standard Model (MSSM); in this
case all the amplitudes should be replaced, \eg
%--
\bqa
\Amp_{\mySM}\lpar \PH \to \PGg \PGg\rpar &\to& 
\Amp_{\MSSM}\lpar \Ph \to \PGg \PGg\rpar
\nl
{}&=& \Amp_{\mySM}\lpar \Ph \to \PGg \PGg\rpar +
      g_{\Ph\PHp\PHm}\,\frac{M^2_{\PW}}{M^2_{\PHpm}}\,A_0\lpar\uptau_{\PHpm}\rpar 
\nl
{}&+& \sum_f\,N^f_c Q^2_f\,
      g_{\Ph\Pft\Pft}\,\frac{M^2_{\PW}}{M^2_{\Pft}}\,A_0\lpar\uptau_{\Pft}\rpar +
      \sum_i\,g_{\Ph\Pcpi\Pcmi}\,
      \frac{M^2_{\PW}}{M^2_{\Pci}}\,A_{\frac{1}{2}}\lpar\uptau_{\Pci}\rpar 
\eqa
%--
with $\uptau_i = \mhs/(4\,M^2_i)$ and
%--
\bq
A_{\frac{1}{2}}(\uptau) = \frac{2}{\uptau^2}\,\Bigl[ \uptau + \lpar \uptau - 1\rpar\,f(\uptau)\Bigr],
\qquad
A_0(\uptau) = - \frac{1}{\uptau^2}\,\Bigl[ \uptau - f(\uptau)\Bigr],
\eq
%--
and
%--
\bq
f(\uptau) = - \frac{1}{4}\,\ln^2\frac{\sqrt{1-\uptau^{-1}} + 1}{\sqrt{1-\uptau^{-1}} - 1}.
\eq
%--
Here $g_{\Ph\PX\PX}$ is the coupling of $\Ph$ to $\PX = \{\PHpm\,,\,\Pft\,,\,\Pcpmi\}$.

Given the number of free parameters in the MSSM that are relevant for Higgs phenomenology,
the present experimental information will clearly not be sufficient to fit the MSSM parameters 
and a further set of Wilson coefficients for $d = 6$ operators. 

An alternative option would be to integrate out the heavy MSSM Higgses (since Buchm\"uller - Wyler 
basis only has a single Higgs field). By squaring the corresponding MSSM interaction Lagrangian 
and contracting the propagators in all possible ways the coefficients will be calculable. 
%--
\subsection{Decoupling \label{acdec}}
%--
In this Section we study the problem of decoupling of high degrees of freedom by considering 
again the decay $\PH \to \PGg \PGg$. To be fully general we assume the existence of heavy 
fermions and scalar that transform according to generic $R_{\Pf}$ and $R_{\PS}$ representations 
of $SU(3)$~\cite{Aglietti:2006tp}. The BSM amplitude is based on couplings
%--
\bq
\PH\Pf\Pf = \frac{1}{2}\,g\,\uplambda_{\Pf}\,\frac{M_{\Pf}}{M_{\PW}},
\qquad
\PH\PS^+\PS^- = g\,\uplambda_{\PS}\,\frac{\mu^2_{\PS}}{M_{\PW}},
\eq
%--
where the $\uplambda$s are numerical coefficients (model dependent) and $\mu_{\PS}$ has the
dimension of a mass. The $\PH\PS^+\PS^-$ vertex follows from the following choice of the
potential:
%--
\bq
V = V_{\mySM} + 2\,\lpar \mPs - \uplambda_{\PS}\,\mu^2_{\PS}\rpar\,
\trac\,\PS^{\dagger} \PS + g^2\,\uplambda_{\PS}\,\frac{\mu^2_{\PS}}{M^2}\,
\lpar \PKdag \PK\rpar\,\trac\,\PS^{\dagger} \PS + \cdots
\eq
%--
where $S= S^a\,T_a$ ($T_a$ are the generators in the $R_{\PS}$ representation) and where the 
trace is over color and $SU(2)$ indices of the field $S$,
%--
\[
\PS = \frac{1}{\srt}\,\left(
\begin{array}{c}
\PS^0_a + i\,\PS^3_a \\
\srt\,i\,\PS^-_a
\end{array}
\right)
\]
%--
\eg $S$ in the $\lpar 8\,,\,2\,,\,\frac{1}{2}\rpar$ of $SU(3) \otimes SU(2) \otimes U(1)$.
The amplitude reads as follows:
%--
\bq
\Amp_{\BSM}\lpar \PH \to \PGg \PGg\rpar =
N^c_{\Pf}\,\uplambda_{\Pf}\,Q^2_{\Pf}\,\Amp_{\Pf} +
N^c_{\PS}\,\uplambda_{\PS}\,\,Q^2_{\PS}\,\frac{\mu^2_{\PS}}{\mPs}\,\Amp_{\PS}.
\eq
%--
where $Q$ is the electric charge of the particle and $N^c$ is the color factor. In the SM we have
%--
\bq
\uplambda_{\Pf} = 1, \quad \uplambda_{\PS} = 0, \quad N^c_{\Pf} = 3 \quad
\mbox{and} \quad R_{\Pf} = 3.
\eq
%--
The amplitudes are
%--
\bq
\Amp_{\Pf} =
\frac{2}{\uptau^2_{\Pf}}\,
\Bigl[ \uptau_{\Pf} + \lpar \uptau_{\Pf} - 1 \rpar\,f\lpar \uptau_{\Pf}\rpar\Bigr]  
\qquad
\Amp_{\PS} =
- \frac{1}{\uptau^2_{\PS}}\,\Bigl[ \uptau_{\PS} - f\lpar \uptau_{\PS}\rpar \Bigr]
\eq
%--
with $\uptau_i = \OMHs/(4\,M^2_i)$. In the limit $M_i \to \infty$ we have
%--
\bq
f\lpar \uptau_i\rpar = \uptau_i + \frac{\uptau^2_i}{3} + {\cal O}\lpar \uptau^2_i\rpar.
\eq
%-- 
The limit $\uptau_i \to 0$ gives
%--
\bq
\Amp_{\BSM}\lpar \PH \to \PGg \PGg\rpar \to
\frac{4}{3}\,N^c_{\Pf}\,\uplambda_{\Pf}\,Q^2_{\Pf} +
\frac{1}{3}\,N^c_{\PS}\,\uplambda_{\PS}\,\,Q^2_{\PS}\,\frac{\mu^2_{\PS}}{\mPs}
\label{coup}
\eq
%--
showing decoupling for $PS$. As stated in \Bref{Aglietti:2006tp} there is decoupling in the
theory when $v = \srt\,M/g \muchless M_{\PS}$; therefore, colored scalars disappear
from the low energy physics as their mass increases (Appelquist-Carazzone 
``decoupling theorem''~\cite{Appelquist:1974tg}).
However, the same is not true for fermions, as shown in \eqn{coup}. We repeat here the
argument of \Bref{Passarino:2011kv}: for a given amplitude involving a massive degree of
freedom (with mass $m$), in the limit $m \to \infty$ we will distinguish {\em decoupling} 
$A \sim 1/m^2$ (or more), {\em screening} $A \to\,$ constant (or $\ln m^2$) and 
{\em enhancement} $A \sim m^2$ (or more). 
Any Feynman diagram contributing to the process has dimension one; however, the total amplitude 
must be proportional to $T^{\mu\nu} = p^{\mu}_2\,p^{\nu}_1 - \spro{p_1}{p_2}\,\delta^{\mu\nu}$ 
because of gauge invariance. For any fermion $\Pf$ the Yukawa coupling is proportional to
$m_{\Pf}/M_{\PW}$ and $T$ has dimension two; therefore, the asymptotic behavior of any diagram 
must be proportional to $T/m_{\Pf}$ when $m_{\Pf} \to \infty$.
The part of the diagram, which is not proportional to $T$, will cancel in the total
because of gauge invariance (all higher powers of $m_{\Pf}$ will go away and this explains the
presence of huge cancellations in the total amplitude).
At LO there is only one Yukawa coupling as in NLO(NNLO) QCD where one add only gluon lines,
so there is screening.

It is worth noting, once again, that electroweak NLO corrections change the scenario: there are 
diagrams with three Yukawa couplings, therefore giving the net  $m^2_{\Pf}$ behavior predicted 
in~\cite{Djouadi:1994ge}, so there is enhancement and, at two-loop level, it goes at most 
with $m^2_{\Pf}$. At the moment, the NLO electroweak corrections for heavy scalar are missing
and no conclusion can be drawn on decoupling at NLO.

In conclusion the decoupling theorem~\cite{Appelquist:1974tg} holds in theories where masses
and couplings are independent. In all theories where masses are generated by spontaneous
symmetry breaking the theorem does not hold in general. Another typical example is given by
the inclusion of a Higgs triplet: if the triplet develops a vacuum expectation value
$v_{\upxi}(v_{\upeta})$ then the $\uprho\,$-parameter deviates from unity at the 
tree-level~\cite{Passarino:1990nu,Aoki:2012yt}  with
%--
\bqa
\uprho_{\myLO} &=& 1 - 2\,\srt\,\myGF\,v^2_{\upxi}, \qquad \mbox{for} \quad Y = 1
\nl
\uprho_{\myLO} &=& 1 + 2\,\srt\,\myGF\,v^2_{\upeta}, \qquad \mbox{for} \quad Y = 0.
\eqa
%--
We will not discuss details of renormalization but one should always remember that whenever 
$\uprho \not= 1$ at tree-level quadratic power-like contribution to $\Delta\uprho$ are absorbed 
by renormalization of the new parameters of the model and $\uprho$ is not a measure of the 
custodial symmetry breaking~\cite{Kanemura:2012rs}. 
Alternatively we could impose custodial symmetry, $v_{\upxi} = v_{\upeta}$, in a model
with both triplets; an example is found in \Bref{Georgi:1985nv} containing
$SU(2)_{\ssL}\,\otimes\,SU(2)_{\ssR}$ multiplets. 

As far as the triplet contribution to $\PH \to \PGg \PGg$ is concerned it is also 
known~\cite{Kanemura:2012rs} that decoupling occurs only for special values of the mixing
angles in the triplet sector.

An important tool in studying decoupling of heavy degrees of freedom is given by
the m-theorem, proved in \Bref{Giavarini:1992xz}: the theorem gives sufficient
conditions for a loop integral to vanish in the large $m\,$-limit. For the one-loop
case it concerns
%--
\bq
I = m^{\alpha}\,\int d^4q\,\frac{P(q)}{\prod_i\,\lpar k^2_i + m^2_i\rpar^{n_i}},
\eq
%--
where
%--
\bq
k_i = q + \sum_{j=1}^N\,\uplambda_{ij}\,p_j,
\qquad
m_i = 0 \; \mbox{or} \; m,
\eq
%--
$P(q)$ is a monomial in the components of $q$, $\{p\}$ are the external momenta and $\alpha$ 
is an arbitrary real number. Let $\upomega$ be the IR degree of $I$ at zero external momenta;
we define
%--
\bq
d = \mbox{dim}\,I,
\qquad
\Upomega = \min\{0\,,\,\upomega\}.
\eq
%--
If $I$ is both UV and IR convergent and $d < \Upomega$ then $I \to 0$ when $m \to \infty$.

In conclusion one should say that BSM Lagrangians can be also classified according to
decoupling. Thus the strategy can be summarized as follows: first, fix benchmark models
to parametrize deviations from the SM, then search for
%--
\[
\mbox{benchmark models} \left\{
\begin{array}{ll}
\quad \in \quad d = 6\,\mbox{operators} & \quad \in \quad \{\Lag_{\dec}\} \\
                                        & \quad \in \quad \{\Lag_{\ndec}\}
\end{array}
\right.
\]
%--
\subsection{Mixing}
%--
There is one assumption in \eqn{FLag} and in its interpretation in terms of ultraviolet 
completions: the absence of mass mixing of the new heavy scalars with the SM Higgs doublet.
Presence of mixings changes the scenario; consider for instance a model with two doublets
and $Y = 1/2$ (THDM), $\upphi_1$ and $\upphi_2$. These doublets are first rotated, with an 
angle $\beta$, to the Georgi-Higgs basis and successively a mixing-angle $\alpha$ diagonalizes 
the mass matrix for the CP-even states, $\Ph$ and $\PH$. 
The SM-like Higgs boson is denoted by $\Ph$ while the VEV of $\PH$ is zero. The couplings of $\Ph$
to SM particles are almost the same of a SM Higgs boson with the same mass (at LO) only if
we assume $\sin\lpar \beta - \alpha\rpar = 1$. Therefore, interpreting large deviations 
in the couplings within a THDM should be done only after relaxing this assumption. 

The case of triplet-like scalars is evem more complex; in the simplest case of a triplet with 
$Y = 1$ there are four mixing angles, all of them entering the coefficients of 
%--
\bq
\frac{1}{\uptau^2_{\PS}}\,\Bigl[ \uptau_{\PS} - f\lpar \uptau_{\PS}\rpar \Bigr]
\eq
%--
in the amplitude for $\Ph \to \PGg \PGg$ (where $\PS = \PH^+, \PH^{++}$) and giving the
couplings $\Ph \PH^+ \PH^-$ and $\Ph \PH^{++} \PH^{--}$, where $\Ph$ is the SM-like Higgs boson.
Only in a very special case, requiring also zero VEV for the triplet, these couplings assume
the simplified form
%--
\bq
c_{\Ph\PH^+\PH^-} = 2\,\frac{M^2_{+}}{v},
\qquad
c_{\Ph\PH^{++}\PH^{--}} = 2\,\frac{M^2_{++}}{v},
\eq
%--
where $v$ is the SM Higgs VEV. Furthermore, decoupling of the charged Higgs partners depends
on the mixing angles and it is the exception not the rule.
%--
\section{Decays into $4\,$-fermions \label{4fdec}}
%--
With a light Higgs boson the decay $\PH \to \PV\PV$ is not open, and one should consider
the full $\PH \to 4\,\Pf$ channel. In order to understand how the calculation can be
organized we start with $\PH \to \PZZ$ where both $\PZ$s are real and on-shell.
%--
\begin{itemize}
\item{\underline{$\PHb \to \PZb \PZb$}}
\end{itemize}
%--
The SM amplitude is
%--
\bq
\Amp^{\mu\nu}_{\mySM} =
- g\,\frac{\OM}{\cths}\,\Bigl\{
\Bigl[ F^{\mySM\,,\,\myLO} + \frac{g^2}{16\,\pi^2}\,F^{\mySM\,,\,\myNLO}_{\ssD}\Bigr]\,
\delta^{\mu\nu} +
\frac{g^2}{16\,\pi^2}\,F^{\mySM\,,\,\myNLO}_{\ssT}\,T^{\mu\nu}
\Bigr\},
\eq
%--
where 
%--
\bq
T^{\mu\nu} = \frac{p^{\nu}_1 p^{\mu}_2}{\spro{p_1}{p_2}} - \delta^{\mu\nu}. 
\eq
%--
We introduce auxiliary coefficients
%--
\bq
A^{\pm}_{\PK} = A^5_{\PK}\,\sth \pm A^4_{\PK}\,\cth,
\qquad
\mybar{A}^0_{\PK} = A^1_{\PK} + A^3_{\PK} + 2\,A_{\dPK}.
\eq
%--
The full amplitude reads as follows
%--
\bq
\Amp^{\mu\nu} = 2^{5/4}\,\myGF^{1/2}\,\lpar
\Amp_{\ssD}\,\delta^{\mu\nu} + \Amp_{\ssT}\,T^{\mu\nu}\rpar, 
\eq
%--
\bqa
\Amp_{\ssD} &=&
   \frac{g_6}{4\,\srt}\,\frac{\OMs}{\cth^3}
   \Bigl[ \lpar 8\,A^3_{\PV}\,\sth\,\cth + A^1_{\PK} - 4\,A_{\dPK}\rpar\,\cth + 
      2\,A^{+}_{\PK}
   \Bigr] 
\nl
{}&-& \frac{\OMs}{\cths} \,F^{\mySM\,,\,\myLO}\, \Bigl[ 1 -
    \frac{g_6}{4\,\srt}
    \,\lpar 8\,A^3_{\PV}\,\sth\,\cth - \mybar{A}^0_{\PK} \rpar \Bigr]
\nl
{}&-& \frac{\myGF \OM^4}{2\,\srt\,\cths\,\pi^2} \, F^{\mySM\,,\,\myNLO}_{\ssD} \,\Bigl[
     1  - \frac{g_6}{4\,\srt}\, \lpar 8\,A^3_{\PV}\,\sth\,\cth 
     - \mybar{A}^0_{\PK} \rpar \Bigr]
\eqa
%---
\bqa
\Amp_{\ssT} &=&
   - \frac{g_6}{\srt}\,\OMHs\,  
     \Bigl[ A^3_{\PV}\,\cth\,\sth - A^2_{\PV}\,\sths - A^1_{\PV}\,\cths - 
     \frac{1}{4}\,A^{-}_{\PK}\,\frac{1}{\cth}\Bigr]
\nl
{}&-& \frac{\myGF \OM^4}{2\,\srt\,\cths\,\pi^2} \, F^{\mySM\,,\,\myNLO}_{\ssT} \,\Bigl[
     1  - \frac{g_6}{4\,\srt}\,\lpar 8\,A^3_{\PV}\,\sth\,\cth 
     - \mybar{A}^0_{\PK} \rpar \Bigr]
\eqa
%--
Following the same strategy we consider the case $\OM_{\PH} < 2\,\OM_{\PZ}$. The process to
consider is then
%--
\begin{itemize}
\item{\underline{$\PHb \to \PZb \PAfb \Pfb$}}
\end{itemize}
%--
which means $\PH \to \PZZ^* \to \PZ \PAf \Pf$. If we work at LO and only include 
local operators proportional to the tree-level $\PH\PZZ$ coupling, then the
SM amplitude is multiplied by a factor
%--
\bq
\Amp^{\PZ}_{\myLO} = 
\Amp^{\mySM}_{\myLO}\,\lpar 1 + \frac{g_6}{4\,\srt}\,\kappa^{\PZ}_{\AC}\rpar
\eq
%--
where the correction w.r.t. the SM is given by
%--
\bq
\kappa^{\PZ}_{\AC} = 
\frac{\OMHs}{\OMs}\,\cths\,\lpar \cth\,A^4_{\PK} - \sth\,A^5_{\PK}\rpar 
- A^1_{\PK} + 4\,A_{\dPK} - 4\,\sth\cth\,A^3_{\PV}
\eq
%--
Therefore we can use (at LO), the SM result and write
%--
\bq
\Gamma\lpar \PH \to \PZZ^*\rpar = \sum_f\,\Gamma\lpar \PH \to \PZZ^* \to 
\PZ \PAf \Pf\rpar = \lpar 1 + \frac{g_6}{2\,\srt}\,\kappa^{\PZ}_{\AC}\rpar\,
\Gamma_{\mySM}\lpar \PH \to \PZZ^*\rpar,
\eq
%--
where the SM partial width is
%--
\bq
\Gamma_{\mySM}\lpar \PH \to \PZZ^*\rpar = 
\frac{\myGF^2 \OMzq}{64\,\pi^3}\,\OM_{\PH}\,F\lpar \frac{\OMzs}{\OMHs}\rpar\,
\lpar 7 - \frac{40}{3}\,\sths + \frac{190}{9}\,\sthq\rpar.
\eq
%--
$F$ is the three-body decay phase-space integral,
%--
\bqa
F(x) &=& \lpar x - 1\rpar\,\lpar \frac{47}{2} x - \frac{13}{2} + \frac{1}{x}\rpar +
\frac{3}{2}\,\lpar 1 - 6 x + 4 x^2\rpar\,\ln x
\nl
{}&+& 3\,\frac{1 - 8 x + 20 x^2}{\sqrt{4\,x -1}}\,
\arccos\lpar \frac{1}{2}\,\frac{3\,x - 1}{x^{3/2}}\rpar.
\eqa
%--
Note that this result cannot be extended beyond LO.
%--
\begin{itemize}
\item{\underline{$\PHb \to \PWb \PAfb \Pfb'$}}
\end{itemize}
%--
Similarly to the previous case we have a correction factor
%--
\bq
\kappa^{\PW}_{\AC} = \frac{\OMHs}{\OMs}\,
A^4_{\PK} - A^1_{\PK} + 2\,A^3_{\PK} + 4\,A_{\dPK}, 
\eq
%--
and the partial decay width can we written as follows:
%--
\bq
\Gamma_{\mySM}\lpar \PH \to \PWW^*\rpar = 
\frac{3\,\myGF^2 \OM^4}{32\,\pi^3}\,\OM_{\PH}\,F\lpar \frac{\OMs}{\OMHs}\rpar.
\eq
%--
Taking the ratio we obtain
%--
\bq
R_{\PZ\PW} = \frac{\Gamma\lpar \PH \to \PZZ^*\rpar}
                  {\Gamma\lpar \PH \to \PWW^*\rpar} =
R^{\mySM}_{\PZ\PW}\,\lpar 1 - \frac{g_6}{2\,\srt}\,r_{\PZ\PW}\rpar
\eq
%--
where the correction factor is
%--
\bq 
r_{\PZ\PW} =  2\,A^3_{\PK} + 4\,\sth\cth\,A^3_{\PV} +
             \frac{\OMHs}{\OMs}\,\sth\,\lpar \sth\,A^4_{\PK} + \cth\,A^5_{\PK}\rpar.
\eq
%--

\begin{itemize}
\item{\underline{$\PHb \to \PZb\PZb \to \PAfb \Pfb \PAfb' \Pfb'$}}
\end{itemize}
%--
However, if we want to deal with the whole process without approximations the final state 
is $4\,$-fermions (say $\Pe\Pe\PGm\PGm$) and the SM amplitude has also non-factorizable 
contributions (\eg pentagons).
%--
\bq
\Amp_{\mySM} = \Amp^{\nu\nu}_{\fact}\lpar p_1, p_2\rpar\,
\Delta_{\mu\alpha}\lpar p_1\rpar\,\Delta_{\nu\beta}\lpar p_2\rpar\,
J^{\alpha}\lpar q_1,k_1\rpar\,J^{\beta}\lpar q_2,k_2\rpar +
\Amp_{\nfact}\lpar p_1, p_2\rpar,
\eq
%--
where $J$ is the fermionic current
%--
\bq
J^{\mu}\lpar q,k\rpar = g\,{\mybar{u}}(q)\,\gamma^{\mu}\,
\lpar v_{\Pf} + a_{\Pf}\,\gamma^5\rpar v(k), \qquad
p = q + k.
\eq
%--
Furthermore, $\Delta^{\mu\nu}(p)$ is the $\PZ$ propagator and $\Amp_{\nfact}$ collects
all diagrams that are not doubly ($\PZ$) resonant. The question is: can we extract informations on
%--
\bq
F^{\mySM}_{\ssD} = F^{\mySM\,,\,\myLO} + \frac{g^2}{16\,\pi^2}\,F^{\mySM\,,\,\myNLO}_{\ssD}
\qquad
F^{\mySM}_{\ssT} =\frac{g^2}{16\,\pi^2}\,F^{\mySM\,,\,\myNLO}_{\ssT},
\eq
%--
or deviations from the two SM structures from the decay $\PH \to 4\,\Pf$?

The form of the $\PZ$ propagator depends on the choice of gauge but, as long as the fermion
current is conserved all differences are irrelevant. 
With the polarization vectors of \appendx{PolV}  one obtains
%--
\bq
\sum_{\uplambda=-1,+1}\,e_{\perp\,\mu}(p,\uplambda)\,e^*_{\perp\,\nu}(p,\uplambda)=
\delta_{\mu\nu} - \frac{p_{\mu} p_{\nu}}{p^2} - e_{\ssL\,\mu}(p)\,e^*_{\ssL\,\nu}(p).
\label{polsum}
\eq
%--
and we can safely replace the $\delta^{\mu\nu}$ in the propagator with a sum over
polarizations, even for off-shell $\PZ$s. Using \eqn{polsum} we replace
%--
\bq
\Delta^{\mu\nu}(p) \to 
\sum_{\uplambda}\,e_{\mu}(p,\uplambda)\,e^*_{\nu}(p,\uplambda)\,\Delta(p^2),
\qquad
\Delta(p^2)= \frac{1}{s + \OMzs},
\eq
%--
with $\OMzs = \OMs/\cths$, $p^2= -s$ and
%--
\bq
e_{\mu}(p,0) = e^{\mu}_{\ssL}(p),
\qquad
e_{\mu}(p,\pm 1) = e^{\mu}_{\perp}(p,\pm 1).
\eq
%--
We introduce the following matrices
%--
\bq
P_{ij} = \Bigl[ \Amp_{\ssD}\,\delta^{\mu\nu} + \Amp_{\ssT}\,T^{\mu\nu}\Bigr]\,
         e_{\mu}(p_1,i)\,e_{\nu}(p_2,j),
\eq
%--
\bq
D_{ij}(p) = \sum_{\spin}\,E_i(p)\,E^{\dagger}_j(p),
\qquad 
E_i(p) = J^{\mu}\lpar q,k\rpar\,e^*_{\nu}(p,i)
\eq
%--  
where $i,j = -1,0,+1$ and $p= q + k$. We obtain
%--
\bqa
\sum_{\spin}\,\bmid \Amp_{\fact}\bmid^2 &=&
\sum_{i j k l}\,P_{ij}\,P^{\dagger}_{kl}\,D_{ik}(p_1)\,D_{jl}(p_2)\,
\bmid \Delta(s_1)\,\Delta(s_2)\bmid^2 = 
\sum_{i j k l}\,A_{ijkl}\,\bmid \Delta(s_1)\,\Delta(s_2)\bmid^2
\nl
{}&=& \Bigl[ \sum_{i}\,A_{iiii} + \sum_{ij}\,A_{ijij} + 
\sum_{\stackrel{k,j \not= i}{l \not= j}}\,A_{ijkl} \Bigr]
\,\bmid \Delta(s_1)\,\Delta(s_2)\bmid^2.
\label{FactA}
\eqa
%--
where $\Amp$ is the matrix element comprising all factorizable contributions, not only the 
SM ones. $A_{iiii}$ gives informations on $\PH$ decaying into two $\PZ$ of the
same helicity ($0,0$ \etc), $A_{ijij}$ on mixed helicities ($0,1$ \etc) while
the third term gives the interference. Therefore
%--
\bq
A_i = A_{iiii}, \qquad A_{ij} = A_{ijij},
\eq
%--
are good candidates to define pseudo-observables. The final step is achieved through the
realization that pseudo-observables are defined in one-point of phase-space and the choice
must respect gauge invariance~\cite{Passarino:2010qk}. The amplitude in \eqn{FactA} has the 
general structure
%--
\bqa
\Amp_{\fact} &=& \sum_{ij}\,a_{ij}\lpar s,s_1,s_2,\,\dots\rpar\,\Delta(s_1)\,\Delta(s_2)
\nl
{}&=& \sum_{ij}\,a_{ij}\lpar \cph,\cpz,\cpz\,\dots\rpar\,\Delta(s_1)\,\Delta(s_2) +
      N\lpar s,s_1,s_2,\,\dots\rpar,
\eqa
%--
where $N$ denotes the remainder of the double expansion around $s_{1,2} = \cpz$, 
$s= -(p_1 + p_2)^2$ and
%--
\bq
\Delta(s) = \frac{1}{s - \cpz},
\eq
%--
$\cph,\cpz$ being the $\PH,\PZ$ complex poles. Therefore, we define pseudo-observables
%--
\bq
\Gamma_i = \int d\Phi_{1 \to 4}\,\sum_{\spin}\,\bmid\,
a_{ii}\lpar \cph,\cpz,\cpz\,\dots\rpar\,\Delta(s_1)\,\Delta(s_2)\,\bmid^2,
\eq
%--
with similar definitions for $\Gamma_{ij}$. Since the problem is extracting pseudo-observables, 
analytic continuation is performed only after integration over all variables but $s_1, s_2$. 
Nevertheless, if one wants to introduce cuts on differential distributions alternative algorithms 
must be introduced, see \Bref{Goria:2011wa}.

The matrices $D, E$ are given by:
%--
\bqa
P_{0\,0} &=& -\frac{1}{2}\,\lpar s_1 s_2\rpar^{-1/2}\,
z_{\PH}\,\Bigl[ \Amp_{\ssD} - 4\,\frac{s_1 s_2}{z_{\PH}^2}\,\Amp_{\ssT} \Bigr],
\nl
P_{+\,+} &=& P_{-\,-} = -\,i\,\lpar \frac{N_{\ssL}}{s_1 s_2 N^1_{\perp} N^2_{\perp}}\rpar^{1/2}\,
\lpar \Amp_{\ssD} - \Amp_{\ssT} \rpar\,
\ep\lpar k_1,k_2,q_1,q_2\rpar, 
\nl
P_{+\,-} &=& P_{-\,+} =
\frac{1}{8}\,\lpar s_1 s_2 N^1_{\perp} N^2_{\perp}\rpar^{-1/2}\,
\lpar \Amp_{\ssD} - \Amp_{\ssT} \rpar\,
\Bigl\{ 2\,s_1 s_2 s_{45} - \Bigl[ s_{46} s_{56} - s_6\,z_{\PH} \Bigr]\,z_{\PH} \Bigr\},
\eqa
%--
where we have introduced
%--
\bq 
s_{ij} = s_i + s_j,
\quad
z_{\PH} = s_{\PH} - s_1 - s_2,
\quad
\ep \lpar k_1,k_2,q_1,q_2\rpar = \ep_{\mu\nu\alpha\beta}\,
k^{\mu}_1\,k^{\nu}_2\,q^{\alpha}_1\,q^{\beta}_2.
\eq
%--  
The elements of the $D\,$-matrix are given by
%--
\bqa
D_{0 0}(p_1) &=& 2\,\lpar V^2_{+} + V^2_{-}\rpar \, 
    \Bigl[ 4 + \frac{1}{N_{\ssL}} \, \lpar 2\,s_1\,s_2 - (s_{34}^2 + s_{56}^2)\rpar\Bigr]
\nl
D_{{-} {-}}(p_1) &=& D_{{+} {+}} =
         \lpar V^2_{+} + V^2_{-}\rpar\,\frac{1}{N^1_{\perp}} \, \Bigl\{
           (s_{34} + s_{56})\,s_{56}
\nl
{}&+& \frac{1}{4}\,\frac{1}{N_{\ssL}} \, \Bigl[
           2\,(s_1^2\,s_2^2 - (s_{34}^2 - 2\,s_1\,s_2)\,s_{56}^2)
          - (s_{34} - s_{56})^2\,s_1\,s_2
          - (s_{34} + s_{56})^2\,s_{56}^2
          \Bigr]\Bigr\}]
\nl
D_{0 {-}}(p_1) &=&
         \frac{i}{\srt N_{\ssL} N^1_{\perp}}\,\lpar V^2_{+} - V^2_{-}\rpar \, 
          \Bigl[ 2\,N_{\ssL} - \frac{1}{2}\,(s_{34} - s_{56})^2 \Bigr]
\nl
{}&+& \frac{1}{\srt N^1_{\perp}}\lpar V^2_{+} + V^2_{-}\rpar \, 
          \Bigl\{  4\,s_{56}
         + \frac{1}{N_{\ssL}} \, \Bigl[
           (s_{34} + s_{56})\,( s_1\,s_2 - s_{56}^2)
          - 2\,(s_{34}^2 - s_1\,s_2)\,s_{56}\Bigr]\Bigr\}
\nl
D_{0 {+}}(p_1) &=&
         \frac{i}{\srt N_{\ssL} N^1_{\perp}}\,\lpar V^2_{+} - V^2_{-}\rpar \, 
          \Bigl[ 2\,N_{\ssL} - \frac{1}{2}\,(s_{34} - s_{56})^2 \Bigr]
\nl
{}&-& \frac{1}{\srt N^1_{\perp}}\lpar V^2_{+} + V^2_{-}\rpar \, 
          \Bigl\{  4\,s_{56}
         + \frac{1}{N_{\ssL}} \, \Bigl[
           (s_{34} + s_{56})\,( s_1\,s_2 - s_{56}^2)
          - 2\,(s_{34}^2 - s_1\,s_2)\,s_{56}\Bigr]\Bigr\}
\nl
D_{{-} {+}}(p_1) &=&
       - \frac{i}{2 N_{\ssL} N^1_{\perp}}\,\lpar V^2_{+} - V^2_{-}\rpar\, 
          (s_{34} - s_{56})\,(s_{56}\,s_{34} - s_1\,s_2)
\nl          
{}&+& \lpar V^2_{+} + V^2_{-}\rpar\,\frac{1}{N^1_{\perp}} \, \Bigl\{
          - (s_{56}^2 + s_1\,s_2)
         + \frac{1}{4}\,\frac{1}{N_{\ssL}} \, \Bigl[
           (s_{34} - s_{56})^2\,s_1\,s_2
          + (s_{34} + s_{56})^2\,s_{56}^2
\nl
{}&-& 2\,(s_1^2\,s_2^2 - (s_{34}^2 - 2\,s_1\,s_2)\,s_{56}^2)
          \Bigr]\Bigr\}
\eqa
%--
and similarly for $D_{i j}(p_2)$.

The result of \eqn{FactA} does not include non-factorizable diagrams. To include them
we will follow the work of \Bref{Bredenstein:2006rh} where standard matrix elements (SME)
are introduced (see Eq. (3.1) and Eq. (3.2) of \Bref{Bredenstein:2006rh}); they are made of
products of
%--
\bq
\Gamma^{i\,,\,\upsigma}_{\mu} =
\frac{1}{2}\,{\mybar{u}}(q_i)\,\gamma_{\mu}\,\lpar 1 + \upsigma\,\gamma^5\rpar v(k_i), 
\qquad
\Gamma^{i\,,\,\upsigma}_{\mu\nu\alpha} =
\frac{1}{2}\,
{\mybar{u}}(q_i)\,\gamma_{\mu}\gamma_{\nu}\gamma_{\alpha}\,
\lpar 1 + \upsigma\,\gamma^5\rpar v(k_i), 
\eq
%--
with $\upsigma= \pm 1$ and $i = 1,2$. For example one has
%--
\bq
\Amp^{12\,,\,\upsigma\uptau} =
\Gamma^{1\,,\,\upsigma\,;\,\mu}\,\Gamma^{2\,,\,\uptau}_{\mu\nu\alpha}\,q^{\nu}_1\,k^{\alpha}_2,
\eq
%--
\etc The non-factorizable amplitude becomes a sum
%--
\bq
\Amp_{\nfact} = \sum_i\,F^{12\,,\,\upsigma\uptau}_i\,\Amp^{12\,,\,\upsigma\uptau}_i,
\eq
%--
where the $F$ are Lorentz invariant form-factors computed up to NLO but excluding those that
are double resonant; the full answer follows by adding this amplitude to $\Amp_{\fact}$.
Note that
%--
\bq
J_{\mu}\lpar q_i,k_i\rpar = g\,\lpar v_{\Pf} + a_{\Pf}\rpar\,\Gamma^{i\,,\,+}_{\mu} +
g\,\lpar v_{\Pf} - a_{\Pf}\rpar\,\Gamma^{i\,,\,-}_{\mu}.
\eq
%--
\section{Double Higgs production \label{dHp}}
%--
A non-zero value of $a_{\Pg}$ fives a contribution also to the $\Pg\Pg\PH\PH$
vertex, contributing to double Higgs production, $\Pg\Pg \to \PH\PH$ (see also 
\Bref{Contino:2012xk}).
%--
\vspace{1.cm}

\fbox{\rule[0.4cm]{0.cm}{1.cm}
\begin{minipage}{0.95\textwidth}
\vspace{0.6cm}
\begin{picture}(0,0)(0,0)
 \SetScale{0.4}
 \SetWidth{1.8}
%--
\Gluon(0,50)(50,0){2}{5}
\Gluon(0,-50)(50,0){2}{5}
\DashLine(50,0)(100,50){3}
\DashLine(50,0)(100,-50){3}
\GCirc(50,0){7}{0.6}
%--
\Text(5,25)[]{$\mu\,a$}
\Text(5,-25)[]{$\nu\,b$}
%--
\end{picture}
%--
\vspace{-1.cm}
%--
\bqa
\Pgb\Pgb\PHb\PHb \quad &{}& \quad 4\,a_{\Pg}\,\myGF\,g_6\,T_{\mu\nu}\,\delta^{a,b}.
\eqa
%--
\vspace{0.5cm}
%--
\end{minipage}}
%--
\vspace{1.5cm}

An additional contribution to double-Higgs production come from the $\PH\PH\PH$ vertex
where $p_1, p_2, p_3$ are the momenta of the outgoing bosons with $p_1 + p_2 + p_3 = 0$.
There are also quartic couplings
%--
\vspace{1.5cm}

\fbox{\rule[0.4cm]{0.cm}{1.cm}
\begin{minipage}{0.95\textwidth}
\vspace{1.cm}
\begin{picture}(0,0)(0,0)
 \SetScale{0.4}
 \SetWidth{1.8}
%--
\DashLine(0,0)(50,0){3}
\DashLine(50,0)(100,50){3}
\DashLine(50,0)(100,-50){3}
\GCirc(50,0){7}{0.6}
%--
\end{picture}
%--
\vspace{-1.cm}
%--
\bqa
\PHb\PHb\PHb \;&{}&\;
  - 3\,\lpar \srt\,\myGF\rpar^{1/2}\,\Bigl\{ 1 + \frac{g_6}{12\,\srt}\,\Bigl[
    3\,A^3_{\PK} + 6\,A_{\dPK} + 32\,\frac{\OMs}{\OMHs}\,A_{\PK}
\nl
{}&-&\,2\,\frac{\sum_{i=1}^3\,p^2_i}{\OMHs}\,
  \mybar{A}^0_{\PK}\Bigr]\Bigr\},
\eqa
%--
\end{minipage}}
%--
\vspace{1.cm}

\fbox{\rule[0.4cm]{0.cm}{1.cm}
\begin{minipage}{0.95\textwidth}
\vspace{0.3cm}
\begin{picture}(0,0)(0,0)
 \SetScale{0.4}
 \SetWidth{1.8}
%--
\DashLine(0,50)(50,0){3}
\DashLine(0,-50)(50,0){3}
\DashLine(50,0)(100,50){3}
\DashLine(50,0)(100,-50){3}
\GCirc(50,0){7}{0.6}
%--
\end{picture}
%--
\vspace{-1.cm}
%--
\bqa
\PHb\PHb\PHb\PHb &{}& 
- 3\,\srt\,\myGF\,\OMHs\,\Bigl[ 1 + \frac{g_6}{2\,\srt}\,\lpar
  A^0_{\PK} + 32\,\frac{\OMs}{\OMHs}\,A_{\PK}\rpar\Bigr]
\eqa
%--
\vspace{0.5cm}
%--
\end{minipage}}

%--
\section{Perturbative unitarity \label{pu}}
%--
In this section we study constraints from perturbative unitarity. With no informations on
the Higgs boson mass there are two different scenarios in 
$\PV_{\ssL} \PV_{\ssL} \to \PV_{\ssL} \PV_{\ssL}$ scattering:
%--
\begin{enumerate}
\item $\mws, \mzs \muchless \mhs \muchless s$

\item $\mws, \mzs \muchless s \muchless \mhs$
\end{enumerate}
%--
Assuming a light Higgs boson we analyze a new option,
%--
\bei
\item $\mws, \mzs, \mhs \muchless s$.
\eei
%--
The SM result iw well-known, given
%--
\bq
\frac{d}{d t}\,\sigma_{\PV_{\ssL} \PV_{\ssL} \to \PV_{\ssL} \PV_{\ssL}} =
\frac{\bmid T(s,t)\bmid^2}{16,\pi\,s^2},
\qquad
T^0_{\myLO} = \frac{1}{16\,\pi\,s}\,\int_{-s}^0\,dt\,T_{\myLO}
\eq
%--
we derive
%--
\bq
T^0_{\myLO}\lpar \PWpL \PWmL \to \PWpL \PWmL\rpar \sim
- \frac{\myGF \mhs}{4\,\srt\,\pi},
\qquad
s \to \infty
\eq
%--
with a critical mass 
%--
\bq 
\bmid T^0_{\myLO}\lpar M_{\PH} = M_c\rpar \bmid = 1,
\qquad
M^2_c= \frac{4}{3}\srt\,\pi\,\myGF^{-1}.
\eq
%--
Anomalous couplings violates perturbative unitarity. However, one has to be
careful in formulating the problem: the region of interest is
%--
\bei
\item $\mws, \mzs, \mhs \muchless s \muchless \Lambda^2$.
\eei
%--
When $s$ approaches $\Lambda^2$ the effective theory must be replaced by the complete
renormalizable, unitary Lagrangian and it makes no sense to study the limit $s \to \infty$
in the effective theory (for a discussion see \Bref{Degrande:2012wf}).
To summarize, anomalous vertices with ad hoc (scale-dependent) form-factors are frequently
used but one should remember that they cannot be put down to an effective Lagrangian.

However, it is well known that heavy degrees of freedom may induce effects of {\em delayed} 
unitatity cancellation in the intermediate region and these effects could easily be 
detectable~\cite{Ahn:1988fx}. Without using the equivalence theorem, we compute
%--
\bq
T^0_{\mySM + \AC} = \frac{1}{16\,\pi\,\uplambda\lpar s,\OMs,\OMs\rpar}\,
\int_{-s+4\,\OMs}^{-t_0 s}\,dt\,
T_{\mySM + \AC}\,\lpar \PWpL \PWmL \to \PWpL \PWmL\rpar,
\eq
%-- 
with a cut $t_0 >> \OMs/s$ to avoid the Coulomb pole. Longitudinal polarization vectors are defined
as follows~\cite{Passarino:1986bw,Passarino:1983bg}
%--
\bqa
e^{\ssL}_{\mu}(p_1) &=&  \frac{2}{\OM s \beta_{\ssM}}\,
\lpar  \spro{p_1}{p_2}\,p_{1\mu} + \OMs\,p_{2\mu}\rpar
\qquad
e^{\ssL}_{\mu}(p_2) =  \frac{2}{\OM s \beta_{\ssM}}\,
\lpar  \spro{p_1}{p_2}\,p_{2\mu} + \OMs\,p_{1\mu}\rpar
\nl
e^{\ssL}_{\mu}(p_3) &=&  \frac{2}{\OM s \beta_{\ssM}}\,
\lpar  \spro{p_3}{p_4}\,p_{3\mu} + \OMs\,p_{4\mu}\rpar
\qquad
e^{\ssL}_{\mu}(p_4) =  \frac{2}{\OM s \beta_{\ssM}}\,
\lpar  \spro{p_3}{p_4}\,p_{4\mu} + \OMs\,p_{3\mu}\rpar,
\eqa
%--
with $\beta^2_{\ssM} = 1 -4\,\OMs/s$. In the limit $\mws, \mzs, \mhs \muchless s 
\muchless \Lambda^2$ we obtain the following
result
%--
\bqa
T^0_{\mySM + \AC} &=&
 - \frac{1}{6}\,\lpar 2 + 5\,t_0 -  t^2_0\rpar\,\lpar 1 - t_0\rpar\,
   \cth\sth\,\lpar 1-2\,\sths\rpar\,\frac{\myGF\,s^2}{\pi\,\OMs}\,A^3_{\PV}\,g_6
\nl
{}&+& \Bigl\{
           \frac{1}{32}\,\lpar 1 - t_0\rpar^2\,
           \lpar A^1_{\PK} + A^3_{\PK} - A_{\dPK} - 6\,\frac{\sth}{\cth}\,A^3_{\PV}\rpar
\nl
{}&+& \Bigl[ \frac{1}{8}\,\lpar 11 + 10\,t_0 - 13\,t^2_0\rpar)
          - 2\,\lpar 1 + 2\,t_0 - 2\,t^2_0\rpar\,\sths\Bigr]\,\cth\sth\,A^3_{\PV}\Bigr\}\,
          \frac{\myGF\,s}{\pi}\,g_6
\nl
{}&+& \frac{3}{16\,\srt}\,\lpar 1 - t_0\rpar^2\,\lpar 
          \frac{\sth}{\cth}\,A^5_{\PK} - A^4_{\PK}\rpar\,
          \frac{\srt\,\myGF^{3/2}\,\OM\,s}{\pi}\,g_6 + \ord{s^0}.
\eqa
%--
As expected the SM part contributes to the constant part while the part proportional
to $g_6$ has positive powers of $s$ (up to power two).
The leading behavior is controlled by the $\Ope^3_{\PV}$ operator.

%--
\section{Conclusions \label{conclu}}
%--
We have described possible deviations from the Standard Model parametrized in terms of
effective $d=6$ operators made of Higgs, gauge and fermion fields, without making the hypothesis 
that the new physics shows up in the Higgs sector. Furthermore, we allow effective operators 
generated at tree level and by loops of heavy particles. 

In this paper we have discussed the implementation of effective Lagrangians with 
emphasis on renormalization. Examples of Lagrangians producing the $d = 6$ operators have been
shown and we have discussed both the decoupling and the non-decoupling scenarios.
In agreement with the work of \Bref{Degrande:2012wf} we have been following the effective field 
theory approach which is cleaner than that of anomalous couplings. An effective field
theory is the low-energy approximation ($E \muchless \Lambda$) to the new physics and it
is only useful up to $E \approx \Lambda$: above $\Lambda$ it should be replaced by a new 
effective theory, parametrizing the low-energy effects at a yet higher scale.

Effective theories should not be considered beyond their UV cutoff, although this is often done
in the literature with the introduction of methods for {\em unitarizing} the model, \eg
form-factors are introduced; this requires specific assumptions and cannot be formulated
in terms of an effective Lagrangian.

There are many scenarios, \eg an interesting one (see \Bref{ArkaniHamed:2012kq}) has no new 
charged fermions and only new bosons. This would unambiguously rule out a large class of BSM 
theories. There are also scenarios with new physics which will be extremely difficult
to distinguish from minimal SM, \eg see \Bref{Dawson:2012di}.
However, the analysis of all possible options should not be done hiding uncertainties 
or the bias from discovering using the minimum $p\,$-value.
Opportunities for precision measurements and BSM sensitivity have been recently described
in \Bref{Mangano:2012mh} and in \Bref{Djouadi:2012rh}.

One final comment is needed: the strategy described in the Introduction amounts to search 
for deviations around a minimum which we assume to be SM. If measured deviations will be large, 
we will face a problem of interpretation: indeed, consider the ratio
%--
\bq
R = \frac{g_{\Ph\PW\PW}}{\ctws\,g_{\Ph\PZ\PZ}},
\eq
%--
where $g_{\Ph\PV\PV}$ is the tree-level coupling of the scalar resonance $\Ph$ to $\PV\PV$;
if $\Ph = \PH$, the SM Higgs boson, then $R = 1$. Assume that $R_{\exp}$ turns out to be close 
to $-1/2$, this will be hard to interpret in terms of a weakly coupled theory and 
it becomes questionable to trust predictions from an effective Lagrangian, based on $d = 6$ 
operators, with Wilson coefficients of that size; to state it differently, $d = 8$ operators and 
insertion of $d = 6$ operators in SM loops are all equally important.
However, some anomalous value of $R_{\exp}$ could very well be close to another weakly coupled 
theory; for instance, the $\Ph_5$ of the Georgi - Machaceck model~\cite{Georgi:1985nv} has 
$R = -1/2$. Starting from the new weakly-coupled Lagrangian will allow us to trust the prediction. 
Of course, one would like to be as model-independent as possible without repeating the fit for
many different starting points; however, there are only very few representations of
$SU(2)_{\ssL}\,\otimes\,SU(2)_{\ssR}$ that respect custodial symmetry, and they should be
included in a more comprehensive analysis.

A recent note by ATLAS Collaboration~\cite{ATLAS-CONF-2012-127}, using data taken in $2011$ and
$2012$, reports that, within the current statistical uncertainties, no significant deviations
from the Standard Model couplings are observed.
%--
\Acknowledgments
%--
We gratefully acknowledge several important discussions with A.~David, M.~Duehrssen, C.~Grojean, 
M.~Spira, G.~Weiglein and the LHC Higgs Cross Section Working Group.
%--
\appendix
%--
\section{Appendix: The ghost Lagrangian}
%--
In this Appendix we give the explicit expression for the Faddeev-Popov ghost Lagrangian.
%--
\bqa
\Lag_{\FP} &=&
        \PAXm\,\partial^2\,\PXm + \PAXp\,\partial^2\,\PXp 
       + \PAYz\,\partial^2\,\PYz + \PAYa\,\partial^2\,\PYa
       - \OMs \, \lpar \PAXp\,\PXp + \PAXm\,\PXm \rpar
       - \OxiZs\,\,\frac{\OMs}{\cths} \, \PAYz\,\PYz
\nl
{}&-& \frac{1}{2}\,g\,\OM\,
         \Bigl[ + \lpar a^3_{\PK} + 2\,a_{\dPK}\rpar\,\ACf \Bigr] \, 
         \lpar \PAXp\,\PXp + \PAXm\,\PXm\rpar\,\POH
\nl
{}&+& \frac{i}{2}\,g\,\OM\,\lpar 1 + a^3_{\PK}\,\ACf \rpar \, 
         \lpar \PAXp\,\PXp - \PAXm\,\PXm\rpar \,\POpz
\nl
{}&-& \frac{1}{2}\,g\,\frac{\OM}{\cths}\,
         \Bigl[1 + \lpar 8\,a^3_{\PV}\,\sth\,\cth + a^3_{\PK} + 2\,a_{\dPK}\rpar\,
         \ACf \Bigr] \, \PAYz\,\PYz\,\POH
\nl
{}&+& \frac{i}{2}\,g\,\frac{\OM}{\cth}\,
         \Bigl[1 + \lpar 4\,a^3_{\PV}\,\cth\,\sth - a^3_{\PK}\rpar \,\ACf 
         \Bigr] \, \lpar \PAYz\,\PXm\,\POpp - \PAYz\,\PXp\,\POpm\rpar
\nl
{}&+& \frac{i}{2}\,g\,\OM\,\Bigl\{
        \frac{\sths - \cths}{\cth} + \Bigl[ 
       4\,a^3_{\PV}\,\sth\,\lpar 1 + 2\,\cths\rpar + a^3_{\PK}\,\frac{(\sths + 3\,\cths}{\cth})
        \Bigr]\,\ACf \Bigr\} \, 
          \lpar \PAXp\,\PYz\,\POpp - \PAXm\,\PYz\,\POpm\rpar
\nl
{}&+& i\,g\,\sth\,
         \Bigl[ 1 + \lpar 4\,a^3_{\PV}\,\sth + a^3_{\PK}\,\cth\rpar \,\ACf \,
         \frac{\cth}{\sths}
         \Bigr]\,\lpar \PAXm\,\pdmu\PXm - \PAXp\,\pdmu\PXp)\,\POA_{\mu} 
         - \lpar \PAXp\,\PXp - \PAXm\,\PXm\rpar \,\pdmu\,\POA_{\mu}\rpar
\nl
{}&+& i\,g\,\sth\,
         \Bigl[ 1 + \lpar 4\,a^3_{\PV}\,\cth\,\sth + a^3_{\PK}\rpar\,\ACf \,    
           \frac{\cths}{\sths}
            \Bigr] \, \Bigl[
            \lpar \PAYa\,\pdmu\PXp - \PAXm\,\pdmu\PYa\rpar \,\PWmmu 
\nl
{}&+& \lpar \PAYa\,\PXp - \PAXm\,\,\PYa\rpar \,\pdmu\,\PWmmu
          - \lpar \PAYa\,\PXm - \PAXp\,\,\PYa\rpar \,\pdmu\,\PWpmu 
          - \lpar \PAYa\,\pdmu\PXm - \PAXp\,\pdmu\PYa\rpar\,\PWpmu\Bigr]
\nl
{}&+& i\,g\,\OM\,\sth\,
         \Bigl[ 1 + \lpar 4\,a^3_{\PV}\,\cth\,\sth + a^3_{\PK}\rpar\,\ACf \,
         \frac{\cths}{\sths}
           \Bigr]\,
         \lpar \PAXm\,\,\PYa\,\POpm - \PAXp\,\,\PYa\,\POpp \rpar
\nl
{}&+& i\,g\,\cth\,
         \Bigl[ 1 - \lpar 8\,a^3_{\PV}\,\cth\,\sth + a^3_{\PK}\rpar\,\ACf \Bigr] \, 
          \Bigl[ 
           \lpar \PAXm\,\pdmu\PXm - \PAXp\,\pdmu\PXp\rpar \,\POZ_{\mu} 
         - \lpar \PAXp\,\PXp - \PAXm\,\PXm\rpar \,\pdmu\,\POZ_{\mu}\Bigr]
\nl
{}&+& i\,g\,\cth\,
        \Bigl[ 1 - \lpar 4\,a^3_{\PV}\,\cth\,\sth + a^3_{\PK}\rpar \,\ACf \Bigr] \, 
         \Bigl[
          \lpar \PAYz\,\pdmu\PXp - \PAXm\,\pdmu\PYz\rpar \,\PWmmu 
        + \lpar \PAYz\,\PXp - \PAXm\,\PYz\rpar \,\pdmu\,\PWmmu
\nl
{}&-& \lpar \PAYz\,\PXm - \PAXp\,\PYz\rpar \,\pdmu\,\PWpmu 
        - \lpar \PAYz\,\pdmu\PXm - \PAXp\,\pdmu\PYz\rpar \,\PWpmu\Bigr]
\label{FPLag}
\eqa
%--
\section{Appendix: Polarization vectors \label{PolV}}
%--
A convenient choice for the polarizations in $\PH \to \PV\PV$ is the following:
%--
\bq
e_{\ssL\,\mu}(p_1) = - N^1_{\ssL}\,\bigl( \spro{p_1}{p_2}\,p_{1\mu} + s_1\,p_{2\,\mu} \bigr),
\qquad
e_{\ssL\,\mu}(p_2) = - N^2_{\ssL}\,\bigl( \spro{p_1}{p_2}\,p_{2\mu} + s_2\,p_{1\,\mu} \bigr),
\eq
%--
where $N_{1,2}$ are the normalizations, $p^2_i = - s_i$ and
%--
\bq
e_{\perp\,\mu}(p_i,\uplambda) = \frac{1}{\srt}\,
\bigl[ n_{\mu}(p_i) + i\,\uplambda\,N_{\mu}(p_i) \bigr],
\quad
N_{\mu}(p_i) = (s_i)^{-1/2}\,\ep_{\mu\alpha\beta\rho}\,
n^{\alpha}(p_i)\,e^{\beta}_{\ssL}(p_i)\,p^{\rho}_i,
\eq
%--
\bq
n_{\mu}(p_1) = i\,N^1_{\perp}\,\ep_{\mu\alpha\beta\rho}\,
k^{\alpha}_1\,p^{\beta}_1\,p^{\rho}_2,
\qquad
n_{\mu}(p_2) = i\,N^2_{\perp}\,\ep_{\mu\alpha\beta\rho}\,
k^{\alpha}_2\,p^{\beta}_2\,p^{\rho}_1.
\eq
%--
With this choice one obtains
%--
\bq
\sum_{\uplambda=-1,+1}\,e_{\perp\,\mu}(p,\uplambda)\,e^*_{\perp\,\nu}(p,\uplambda)=
\delta_{\mu\nu} - \frac{p_{\mu} p_{\nu}}{p^2} - e_{\ssL\,\mu}(p)\,e^*_{\ssL\,\nu}(p).
\eq
%--
Using $P \to p_1 + p2_2 \to q_1 + k_1 + q_2 + k_2$ we define 
%--
\bq
\spro{p_1}{p_1} = -s_1 \quad
\spro{p_2}{p_2} = -s_2 \quad
\spro{p_1}{p_2} = \frac{1}{2}\,\lpar s_1+s_2-s_{\PH}\rpar
\eq
%--
\bq
\spro{q_1}{q_1} = 0 \quad
\spro{q_2}{q_2} = 0 \quad
\spro{k_1}{k_1} = 0 \quad
\spro{k_2}{k_2} = 0 \quad
\eq
%--
\bq
\spro{q_1}{k_1} = -\frac{1}{2}\,s_1 \quad
\spro{q_1}{q_2} = -\frac{1}{2}\,s_3 \quad
\spro{q_1}{k_2} = -\frac{1}{2}\,s_4 \quad
\eq
%--
\bq
\spro{k_1}{q_2} = -\frac{1}{2}\,s_5 \quad
\spro{k_1}{k_2} = -\frac{1}{2}\,s_6 \quad
\spro{q_2}{k_2} = -\frac{1}{2}\,s_2,
\eq
%--
where we allow for an off-shell Higgs boson, $P^2= -s_{\PH}$. We derive
%--
\bq
N^i_{\ssL} = \lpar s_i\,N_{\ssL} \rpar^{-1/2},
\qquad
N^2_{\ssL} = \frac{1}{4}\,\uplambda\lpar s_{\PH}, s_1, s_2\rpar,
\eq
%--
where $\uplambda$ is the K\"allen function. Furthermore,
%--
\bqa
N^{-2}_{1\perp} &=& \frac{1}{4}\,\Bigl[ \lpar s_5 + s_6\rpar\,
                                        \lpar s_3 + s_4\rpar - s_1\,s_2 \Bigr],
\nl
N^{-2}_{2\perp} &=& \frac{1}{4}\,\Bigl[ \lpar s_4 + s_6\rpar\,
                                        \lpar s_3 + s_5\rpar - s_1\,s_2 \Bigr].
\eqa
%--
\clearpage
%--
\bibliographystyle{atlasnote}
\bibliography{HLS}{}

%===
\end{document}